\documentclass[conference]{IEEEtran}
\usepackage{cite}
\usepackage{amsmath,amssymb,amsfonts}
\usepackage{algorithmic}
\usepackage{graphicx}
\usepackage{textcomp}
\usepackage{xcolor}
\usepackage[hyphens]{url}

\def\BibTeX{{\rm B\kern-.05em{\sc i\kern-.025em b}\kern-.08em
    T\kern-.1667em\lower.7ex\hbox{E}\kern-.125emX}}

\usepackage[bookmarks=true,breaklinks=true,letterpaper=true,colorlinks,linkcolor=black,citecolor=blue,urlcolor=black]{hyperref}

\usepackage{soul}
\usepackage{enumitem}
\usepackage{booktabs}
\usepackage{float}
\usepackage{subfig}
\usepackage{graphicx}
\usepackage[export]{adjustbox}
\usepackage{array}
\newcolumntype{P}[1]{>{\centering\arraybackslash}p{#1}}
\usepackage{changepage}
\usepackage{comment}

\newcommand{\arch}{Cloak}
\newcommand{\archTextit}{\textit{Cloak}}

\title{A Method for Hiding the Increased Non-Volatile Cache Read Latency} 
      
\usepackage[blocks]{authblk}

\author{Apostolos Kokolis}
\affil{University of Illinois, Urbana-Champaign\\
kokolis2@illinois.edu
}
\author{Namrata Mantri}
\affil{University of Illinois, Urbana-Champaign\\
nmantri2@illinois.edu
\vspace{5pt}
}
\author{Shrikanth Ganapathy\textsuperscript{\textsection}}
\affil{Rivos Inc.\\
shrikanth.ganapathy@gmail.com
}

\author{Josep Torrellas}
\affil{University of Illinois, Urbana-Champaign\\
torrella@illinois.edu
}
\author{John Kalamatianos}
\affil{AMD Inc.\\
john.kalamatianos@amd.com
}

\begin{document}
\maketitle
\begingroup\renewcommand\thefootnote{\textsection}
\footnotetext{Work performed while employed at AMD Inc.}
\endgroup
\thispagestyle{plain}
\pagestyle{plain}


\begin{abstract}
The increased memory demands of workloads is putting high pressure on Last Level Caches
(LLCs). Unfortunately, there is limited opportunity to increase the capacity of LLCs due to the area and power requirements of the underlying SRAM technology. 
Interestingly, emerging Non-Volatile Memory (NVM) technologies promise a feasible alternative to SRAM for LLCs due to their higher area density.
However, NVMs have substantially higher read and write latencies, which
offset  their area density benefit.
Although researchers have proposed methods to tolerate NVM's increased write latency, little emphasis has been placed on reducing the critical NVM read latency.

To address this problem, this paper proposes {\em \arch{}}.
\arch{} exploits data reuse in the LLC at the page level, to hide NVM read latency.
Specifically, on certain L1 TLB misses to a page, \arch{} transfers LLC-resident data belonging to the page from the LLC NVM array to a set of small SRAM Page Buffers that will service subsequent requests to this page. 
Further, to enable the 
high-bandwidth, low-latency transfer of lines of a page to the page buffers,
\arch{} uses an LLC layout that accelerates the discovery of LLC-resident cache lines from the page.
We evaluate \arch{} with full-system simulations of a 4-core processor across 14 workloads. We find that, on average, \arch{} outperforms an SRAM LLC by 23.8\% and
an NVM-only LLC by 8.9\%---in both cases, 
with negligible additional area.
Further, \arch{}'s
\(ED^2\) 
is 39.9\% and 17.5\% lower, respectively, than these designs.
\end{abstract}

\vspace{-1mm}
\section{Introduction}

The popularity of
data intensive workloads, such as HPC 
applications and web servers, cloud applications and databases, has intensified capacity pressure on
Last-Level Caches (LLCs). While much larger LLCs are desired, SRAM technology 
suffers from high area overhead (exacerbated by the 
increasing manufacturing cost at leading edge technologies \cite{semieng3nm,wccftech}), substantial leakage power, and scalability 
problems~\cite{tec_comparison}. 

For these reasons, researchers have examined alternative memory technologies,
such as eDRAM and Non-Volatile Memory (NVM). In particular, NVM technologies such as PCM~\cite{pcm} and, especially, STT-RAM~\cite{stt_ram} are promising candidates to replace SRAM in LLCs. Compared to SRAM, STT-RAM offers higher density and
lower leakage power~\cite{stt_ram_leakage}. Unlike eDRAM,
NVM is compatible with current logic and SRAM, and can be easily integrated 
in the same die. Further, compared to eDRAM, NVM offers lower complexity 
(no refresh, activate, or precharge operations), 
comparable read access time, and improved power-efficiency due to its ability to power gate without losing state.

However, NVMs have two main shortcomings over SRAM, namely, higher 
latency for both read and write operations, and a higher dynamic energy consumption 
per access. Moreover, read and write latencies in NVMs change based on the targeted lifetime endurance (wear-out) of the device. Therefore, replacing an SRAM LLC with an NVM one becomes a trade-off between latency, capacity, reliability and energy consumption.
In this paper, we focus on mitigating the long NVM read latency for highly-reliable NVM caches.

Table~\ref{table:latencies} compares the characteristics of SRAM and STT-RAM cells. We can see that STT-RAM cells are $\sim$4x smaller in area, while their read and write latencies are 10--30x and
25--100x, respectively, higher than SRAM's. The table does not
 include the energy and power numbers because the literature provides wide ranges of values, dependent on implementation and manufacturing technology \cite{ read_barrier, hybrid_cache, oap, adaptive_placement, dasca}. Specifically, STT-RAM's leakage power is 0.15--0.48x that of SRAM's, and its dynamic access energy is 0.8--2.5x higher than SRAM's for reads and 1.5--15x higher for writes.

\begin{table}[h!]
\vspace{-1.5mm}
\begin{center}
\fontsize{8}{9}\selectfont 
 \begin{tabular}{||l|c|c||} 
 \hline \hline
 Characteristic & SRAM & STT-RAM \\ 
 & \cite{7071813,8081898,karl,6903378} & \cite{tapeout1,tapeout2,tapeout3,tapeout4, sakhare} \\
 &
 & \cite{tapeout8,7409762,tapeout5, tapeout6,tapeout7} \\
 [0.5ex] 
 \hline\hline
  Area (\(F^{2}\)) & 70-150 & 15-40\\ 
 Read Latency (ns) & 0.3 & 3 - 10 \\
 Write Latency (ns) & 0.3 & 8 - 30 \\
 \hline  \hline
\end{tabular}
\vspace{-1.2mm}
\caption{Comparing SRAM and STT-RAM characteristics.}
\vspace{-4.mm}
\label{table:latencies}
\end{center}
\end{table}

Prior work has tried to overcome NVM's problems of higher latency---primarily write latency---and dynamic power using solutions spanning the device, circuit, and architecture levels.
At the device and circuit levels, the write access latency 
(primarily) can be
reduced by sacrificing the retention time and
non-volatility of the STT-RAM cells \cite{relax_non_volatility, cache_revive, multi_retention, amnestic_cache}. Also, the transistor size 
can be adjusted for faster write operation \cite{tapeout4}, at the cost of higher power, lower density, and lower reliability \cite{stt_ram_leakage}.
Such approaches limit the full potential of NVM caches and do not solve the increased read latency problem.
Indeed, degrading NVM characteristics to reduce the write latency introduces the need for periodic refresh, which hurts design complexity and energy consumption, and hinders 
non-volatility \cite{memory_persistency, delegated_persist}. 
Additionally, adjusting the NVM cell size to reduce 
write latency
limits NVM capacity and introduces higher error rates~\cite{stt_ram_leakage, stt_ram_variations}.

At the architecture level, the most popular solutions to address NVM's higher latency and dynamic power combine smaller SRAM caches with larger STT-RAM caches in a hybrid cache hierarchy \cite{hybrid_cache, hybrid_cache2, adaptive_placement, novel_3d}. 
Unfortunately, these solutions focus mostly on write 
latency (not the focus of this paper) or
use inclusive caches (much less used today). 
In addition, they use a considerable amount of SRAM memory, 
plus complex logic to decide which cache lines to
swap between SRAM and STT-RAM.
As a result, they limit the area savings 
from NVM and increase the energy consumption. 

In terms of access latencies,
several proposals mitigate the performance impact of long NVM write 
latencies~\cite{density_tradeoffs, selective_slow_writes, oap, lap, oscar, dasca}. 
Interestingly, little emphasis has been placed on mitigating the NVM read latency, based on the common assumption that the SRAM and STT-RAM read latencies are similar. However, measurements on fabricated STT-RAM caches and observations by industry vendors show a significant difference in
read latency between STT-RAM~\cite{read_barrier, stt_llc, hybrid_cache, tapeout1, tapeout2,tapeout3,tapeout4,tapeout5,tapeout6,tapeout7,tapeout8, 7409762} and SRAM~\cite{7071813,8081898,karl}. Specifically, as shown in 
Table~\ref{table:latencies}, 
FinFET-based 6T SRAM arrays perform read operations with a 300ps
latency~\cite{karl}, while the fastest STT-RAMs can only attain
3ns latencies at best~\cite{sakhare}. 

To take advantage of NVM for LLCs, we need a 
low-cost architectural solution that can tolerate the higher read latency of STT-RAM without sacrificing capacity, reliability, or non-volatility. 
This paper proposes such an architectural solution, which we call~\archTextit{}. 
\arch{} exploits data reuse at the level of pages
in the LLC to hide NVM read latency.
Specifically, on certain L1 TLB misses to a page, the hardware transfers LLC-resident lines of the page from the LLC NVM array to a set of small SRAM Page Buffers. Such buffers will service future requests to this page. To enable the low latency detection and high-bandwidth transfer of lines of a page to the page buffers, 
\arch{} uses an LLC layout that accelerates the discovery of LLC-resident cache lines from the page.
Further, we develop an adaptive replacement policy for the Page Buffers to increase their utilization and achieve better performance and energy consumption.

We evaluate \arch{} with full-system simulations of a four-core processor running 14 workloads. On average, \arch{} outperforms an SRAM LLC by 23.8\% and
an NVM-only LLC by 8.9\%---in both cases, with negligible additional area. 
Furthermore, \arch{} improves the \(ED^2\) metric by 39.9\% and 17.5\%, respectively.

\vspace{-0.6mm}
\section{Background}
\vspace{-0.1mm}
\subsection{STT-RAM Limitations and Opportunities}

STT-RAM has emerged as a promising candidate to replace SRAM
in LLCs~\cite{stt_llc, hybrid_cache, novel_3d, tapeout9} 
because it provides higher density and lower leakage than SRAM.
However, the viability of STT-RAM is inhibited by higher read 
and write access latencies, and by higher dynamic energy than SRAM.
This is because STT-RAM requires a high thermal barrier, 
which in turn increases the switching current of the device 
and the cell access latency.
The high thermal barrier is a consequence of the requirement 
for a large retention period, typically expected of non-volatile
memory cells.

Past research has exploited a trade-off that exists between 
retention time and write access latency, to design cells 
whose write access latency is tolerable for practical on-chip integration~\cite{cache_revive, relax_non_volatility, multi_retention, amnestic_cache}. 
Such techniques help lower the 
device write time to 8--30ns, as shown in Table~\ref{table:latencies}.
These proposals focus on reducing write latency because 
 read latencies tend to be smaller, namely
3--10ns.

However, one major caveat is that the STT-RAM array latency cannot be {\em   pipelined}. During the cell  
access latencies reported in Table~\ref{table:latencies},
the array
is blocked from servicing other requests. 
As a result, sustaining high throughput is challenging.
In contrast, SRAM array accesses are pipelined, and so data can be moved to/from the cache array every cycle. This leads to both higher throughput 
and guaranteed maximum latency for accesses.

STT-RAM write latency is constrained by bit-level error guarantees to ensure reliable operation across the lifetime of the chip.
In our case, where the STT-RAM LLC is 16MB per slice (Table~\ref{table:architecture}), the Bit Error Rate (BER) of the STT-RAM cell needs to be lower than 10$^{-10}$ to ensure a 99.99999\% yield with SECDED ECC.
This is based on the assumption that a cache line is fetched from a single STT-RAM macro of 2MB.
Based on results from prototype devices\cite{tapeout4, tapeout5, 8993474}, STT-RAMs with such error rate guarantees can only achieve a bitcell write latency of $\approx$8ns and a read latency of $\approx$3.2ns.
Furthermore, recent innovations in the quality of magnesium oxide (MgO), which acts as the dielectric material,
pave the way for high endurance STT-RAM cells in future designs \cite{9108144}.
As a result, the literature reports that STT-RAM is a competitive
alternative to SRAM for L3 caches. It can have an endurance in the order of 10$^{12}$ to 10$^{13}$ write cycles \cite{hybrid_cache2, density_tradeoffs, 9108144, oscar},
especially under normal temperature environments as the one we target \cite{9062955}.

Overall, given the reasonable tradeoffs between STT-RAM write latency and BER, the most pressing problem is to mitigate the impact of the high read latency of large STT-RAM LLCs operating in ambient conditions. This is our focus.

\vspace{-0mm}
\subsection{NVM Cache as an SRAM Replacement} 
\vspace{-0.1mm}

Prior work on NVM caches has focused on mitigating the effect of 
long-latency write operations~\cite{density_tradeoffs, selective_slow_writes, oap, lap, oscar, dasca}. The work
can be categorized into three groups: methods to reduce, stall or bypass writes to NVM caches, NVM cell optimizations, 
and hybrid SRAM/NVM caches.

Solutions in the first group identify write contention in the NVM cache that can stall latency-critical reads. Then, they try to take writes off the critical path of subsequent read requests~\cite{density_tradeoffs, oap, selective_slow_writes, dasca}. These techniques assume the same read latency for NVM and SRAM accesses.

Proposals that optimize NVM cells improve NVM cache write performance
at the expense of retention time and area~\cite{multi_retention, cache_revive, relax_non_volatility, tapeout4}.
These optimizations are not trivial, given the trade-offs between access latency, area, and retention time of NVMs \cite{stt_llc, stt_ram, stt_ram_variations}. In addition, decreasing the retention time of NVMs introduces refreshes, similar to DRAM, which increase energy consumption and complicate the design. Moreover, it limits the capacity of NVMs and introduces higher error rates. 
Importantly, it does not address the problem of the non-pipelined and high read latency.

Hybrid cache proposals \cite{adaptive_placement, hybrid_cache, hybrid_cache2, novel_3d} split a cache into an NVM and an SRAM portion, typically by partitioning a cache set into SRAM and NVM ways. 
These proposals monitor address reuse to place data in the SRAM or NVM parts of the cache.

For energy efficiency, the SRAM and NVM portions of the cache can be powered on/off according to the cache line usage \cite{hybrid_cache2}. This approach leads to energy savings at the expense of 
performance degradation and increased circuit complexity.

For performance, some techniques \cite{adaptive_placement, hybrid_cache, novel_3d} place the critical data in the SRAM portion of the cache, and the non-critical data in the NVM portion. 
However, these techniques have several shortcomings. 
First, they dedicate a large portion of cache capacity to SRAM, therefore reducing the density benefit of NVM and increasing leakage power. 
The area overhead of the SRAM portion  is 25--100\%, assuming a 4:1 density between NVM and SRAM \cite{novel_3d,adaptive_placement, hybrid_cache}.
Second, they need large structures of several KBs to accurately monitor cache line activity.
Third, they must migrate data between SRAM and NVM, which further increases the number of writes to NVM, the energy consumption, and the area overhead. 
Fourth, in exclusive and victim LLCs, it is hard to monitor the reuse of individual addresses because LLC hits result in promoting lines to faster levels of the cache hierarchy.
As a result, the overhead and complexity of recording data reuse increases. 
Finally, these techniques do not consider that the non-pipelined nature
of NVM accesses introduces high overhead to line migration.

\vspace{-1mm}
\section{Motivation}\label{motivation}
\vspace{-0.2mm}

\begin{figure}[h]
  \centering
  \begin{center}
      \vspace{-3.8mm}
    \hbox{\hspace{-0.4em}\includegraphics[width=1\linewidth]{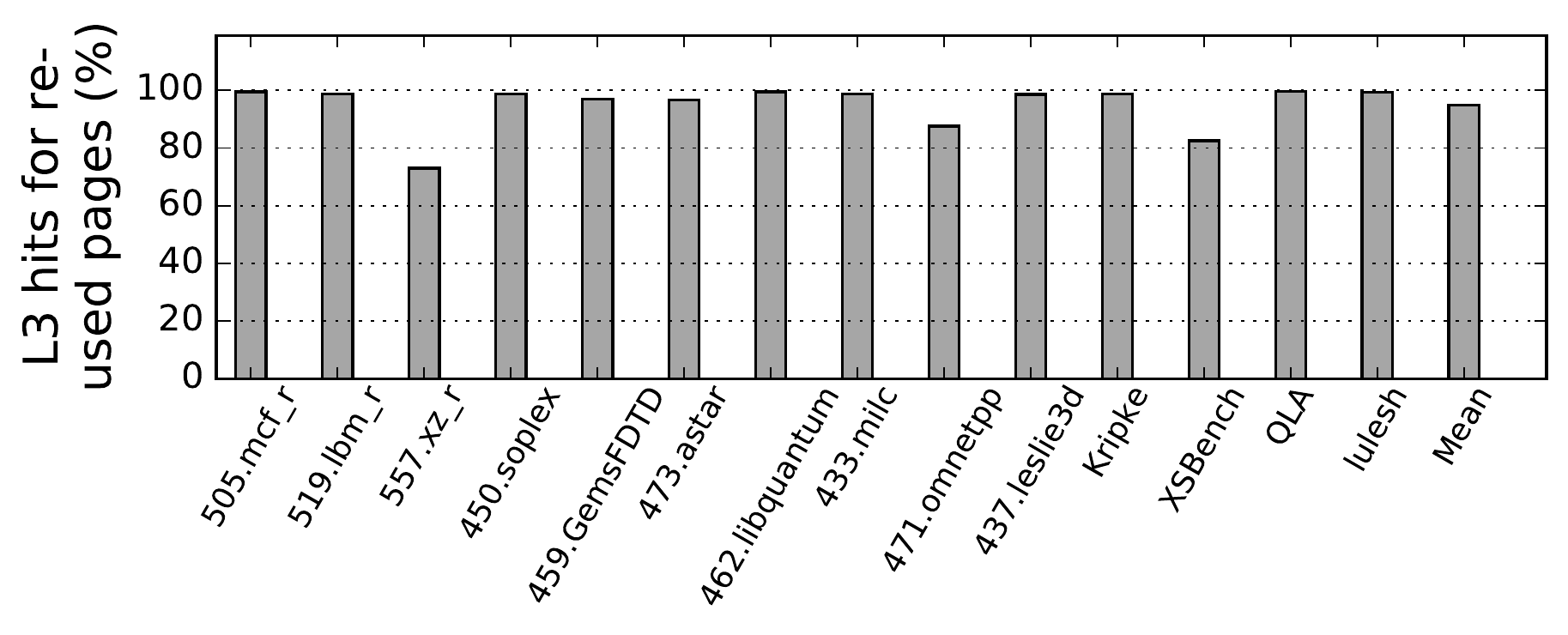}}
    \vspace{-3.1mm}
  \caption{Percentage of L3 hits that originate from accesses to pages that were re-filled into the L1 TLB.}
  \label{fig:reused_hits}
  \end{center}
  \vspace{-7.5mm}
\end{figure}

\begin{figure}[h]
  \centering
  \begin{center}
      \vspace{-0mm}
    \hbox{\hspace{0em}\includegraphics[width=1\linewidth]{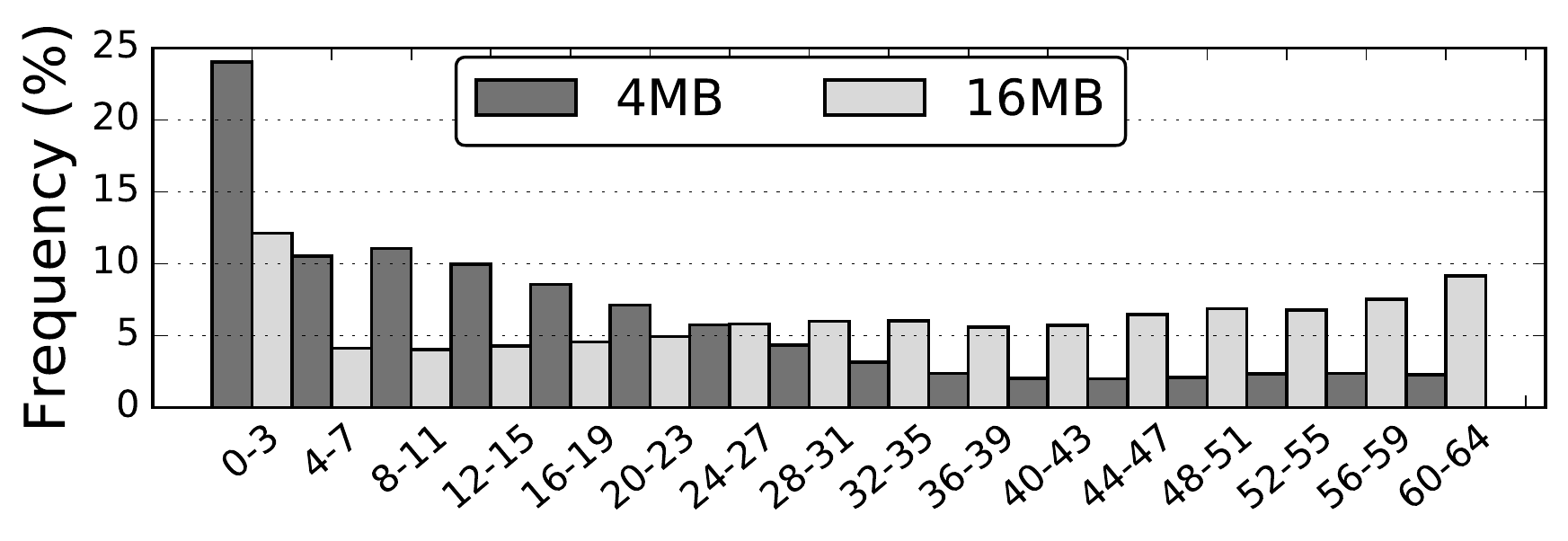}}
        \vspace{-2.6mm}
  \caption{Frequency of the number of L3-resident cache lines from 
  a 4KB page that is re-filled into the L1 TLB.}
  \label{fig:l3resident}
  \end{center}
  \vspace{-8.mm}
\end{figure}

To increase the cache capacity in multi-cores, designers are organizing LLCs as victim caches. We would like to use NVM in the LLC,
to enable high capacity caching with high retention time and low error rates. To this end, in 
this paper, we observe that an L1 TLB miss on a page {\em that was 
already referenced in the past} is a good hint that some LLC-resident cache lines of the page will be reused again soon. Consequently, we would like to identify
such TLB refills and bring likely-to-reuse lines from the LLC into a small SRAM 
structure.

To support this insight, we model a 3-level cache hierarchy with a 
16MB L3 and 4KB pages. Figure~\ref{fig:reused_hits} shows the percentage of L3 hits that originate from pages whose translation was once in the L1 data TLB, was evicted, and then later was re-filled back into the L1 TLB. 
We observe that such percentage is 
94.9\%.
This data implies that, upon an L1 TLB page re-fill, there should be many requests to this page that hit in the LLC.

Figure \ref{fig:l3resident} shows the number of L3-resident cache lines belonging to a 4KB page that is being re-filled into the L1 TLB for different L3
cache sizes. Such number can be from 0 to 64 lines. We see that, while in most
cases we have 0--3 lines, there is a long tail of up to 60-64 resident lines.
In addition, as we scale the L3 size from 4MB to 16MB per core, the number of such resident cache lines increases.
Therefore, we conclude that the number of requests hitting in the L3 cache and originating from a TLB refilled page likely rises with the L3 size.
\arch{} builds on these observations to architect a solution that hides the increased read latency of NVM caches and increases overall throughput.
\vspace{-0.5mm}
\section{Design Overview of CLOAK} 
\vspace{-2.2mm}

\begin{figure}[h]
  \centering
  \begin{center}
      \vspace{-2mm}
    \hbox{\hspace{1em}\includegraphics[width=0.85\linewidth]{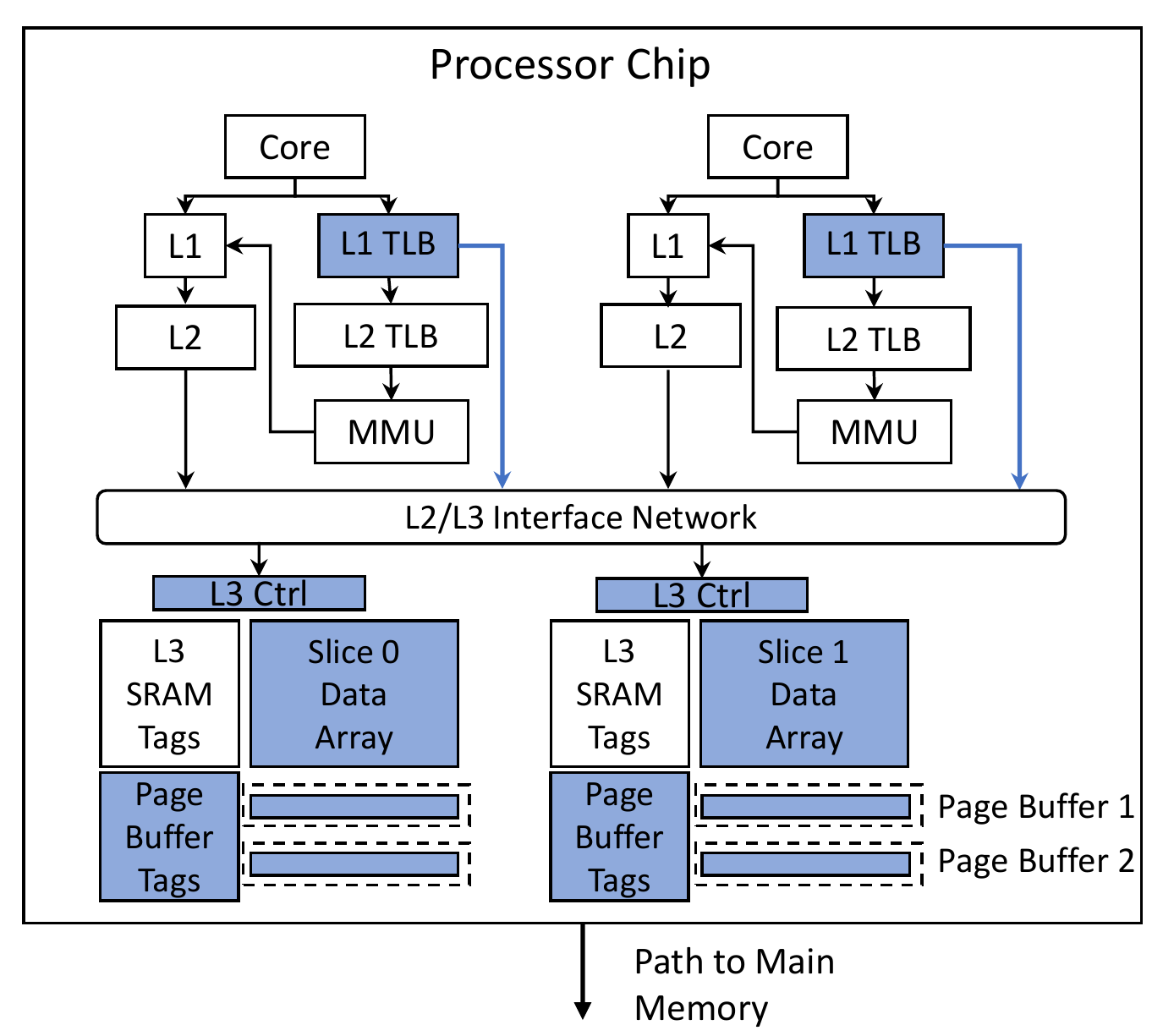}}
    \vspace{-2.2mm}
  \caption{\arch\ architecture, with the new or modified hardware structures shaded,
and the added connections in lighter color.}
  \label{fig:design_overview}
  \end{center}
 \vspace{-11mm}
\end{figure}

\subsection{Main Idea}

\arch{} is a hardware mechanism that takes advantage of certain L1 TLB misses to exploit data re-use in large NVM LLC caches and hide NVM read latency. 
The NVM LLC cache is augmented with small SRAM buffers, which we call \textit{Page Buffers (PB)}. PBs hold data transferred from the NVM LLC main array. Each PB can hold a copy of LLC-resident cache lines originating from the same page.
To trigger the copy of data into a PB, \arch{} leverages the L1 data TLBs.
When a miss in the L1 TLB occurs for a previously-accessed page, hardware communicates this information to the LLC, which finds all  LLC-resident lines from this page and copies them to the PBs.

The lines that were previously accessed from the page have a high chance of being accessed again when the page is re-filled in the L1 TLB---due to temporal locality within a page.
\arch{} tracks page-level reuse via address translation hardware to infer future references to the NVM cache.
To facilitate the retrieval of a page's cache lines from the LLC, we introduce a new LLC data layout that places the cache lines of a given page in the same physical cache row.

Figure \ref{fig:design_overview} shows the architecture of \arch{}, where the new or modified hardware structures are shaded, and the added connections between the L1 TLBs and the L3 Controllers are shown in lighter color. In this section, we describe the overall operation of \arch{} to fetch data to the PBs and service memory requests. Subsequently, Section \ref{implementation} discusses the architectural details of our design.

\subsection{\arch{} Overview} \label{overview}

\begin{figure}[h]
  \centering
  \begin{center}
    \vspace{-2.2mm}
    \hbox{\hspace{0.4em}\includegraphics[width=1\linewidth]{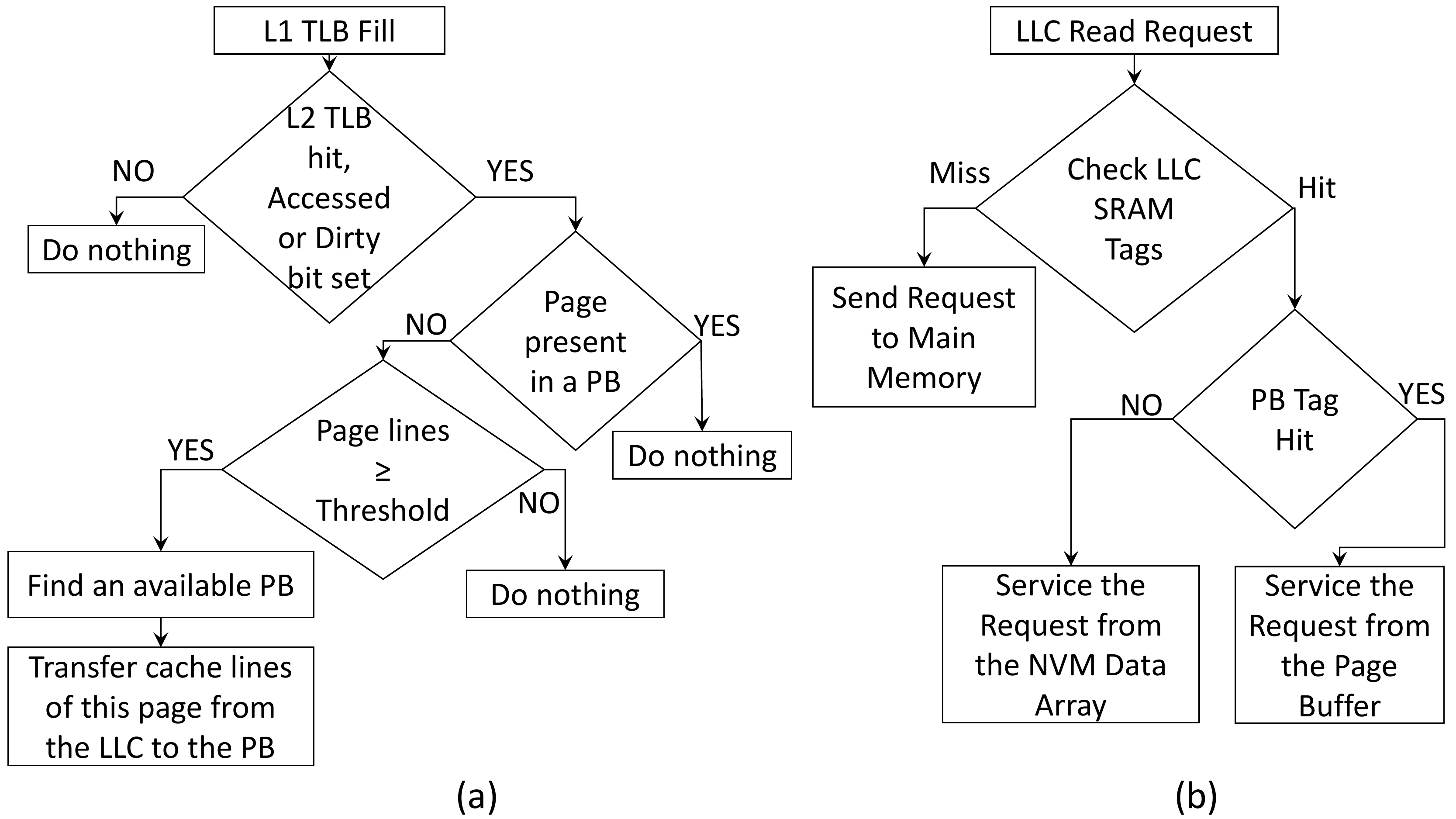}}
    \vspace{-1.5mm}
  \caption{Control flow diagrams: (a) promotion of lines from a page to a PB, and (b) servicing a read request to the LLC.}
  \label{fig:control_flow}
  \end{center}
   \vspace{-8.mm}
\end{figure}

The core operations of \arch{} consist of data movement from the NVM data array to the PBs on a TLB signal, and servicing of subsequent requests either from the PBs or the NVM cache array. Figure \ref{fig:control_flow} shows the control flow diagrams of these actions.

\subsubsection{TLB triggered Page Buffer transfer}

PBs are small SRAM-based cache structures which act as fast access buffers to the NVM LLC. 
Each PB can hold a subset of the NVM-resident cache lines of a given page.
Promoting NVM-resident lines from a page into a PB can hide the read latency of an LLC access, by exploiting intra-page spatial locality~\cite{cache_locality} and by reducing the number of accesses to the NVM data array.

The promotion of a page's cache lines to one of the PBs 
is depicted in Figure~\ref{fig:control_flow}a.
When an L1 TLB miss occurs, the PTE for the page is fetched and 
\arch{} determines whether this page was previously referenced.
To identify whether the page was referenced in the past, \arch{} 
checks if the page is in the L2 TLB or, if it is not, if either the \textit{Accessed} or \textit{Dirty} bits of its PTE~\cite{amd_manual} is set. 
A set Accessed bit indicates that the page  was accessed in the past. This bit is set by the hardware when the page is first read or written, and is only reset by the OS to track the frequency of accesses to the page. A set Dirty bit indicates that the page was written, and hence, was referenced before. The Dirty bit is set by the processor the first time that the page is written to, and is only cleared by software. 

If \arch{} finds that the page  was used in the past, it sends a signal to the LLC controller containing the physical address of the request that caused the TLB refill.
The LLC controller first determines if a PB already contains lines from this page. 
If not, \arch{} decides whether to transfer the page to a PB and, if so, which PB to use.
To decide whether to transfer the page's lines, \arch{} checks the LLC tags to
calculate the population of NVM-resident lines of this page.
A promotion to a PB occurs only if the population size
exceeds a programmable threshold, so that the cost of fetching all the 
lines to a PB can be amortized across the expected number of PB hits.
Then, \arch{} selects an available PB to promote the page's cache lines according to the PB replacement policy (Section \ref{replacement_policy}).

\subsubsection{Servicing requests to the LLC}

In \arch{}, a hit in the LLC can obtain the data from the NVM cache array or 
from a PB.
The PB contents are kept coherent with the NVM cache. 
Thus, writes to the LLC (e.g., due to an L2 eviction) also check the
PBs and, on a hit, update both the NVM data array and the PB.
An acknowledgement, signaling completion of the write request, is sent to the L2 when the request is buffered in the LLC queues. It does not wait until
when the request updates the 
\mbox{NVM data array and PB.}

Figure~\ref{fig:control_flow}b shows servicing a read request to the LLC.
The LLC controller checks the Physical Page Numbers (PPNs) in the PB Tags
and in the LLC SRAM tags in parallel, to determine a hit or a miss. 
If the LLC tag check determines that there is an LLC miss, the request is forwarded to main memory.
If the LLC tag check determines an LLC hit and the PB tag check indicates a PB hit, the request is serviced from the PBs otherwise it is serviced from the NVM cache.

In \arch{}, an LLC NVM hit can be serviced in parallel with a PB hit to a different address.
This does not increase the peak LLC bandwidth because data coming from the PB and from the NVM data array share a single bus to the L2 cache. However, in-flight read accesses to the slow, non-pipelined NVM data array do not block younger reads to the PBs.

\section{Cloak Implementation} \label{implementation}

\subsection{Data Layout}\label{llc_layout}
\vspace{-0mm}

Transferring the lines of a page from the LLC to a PB
requires finding all NVM-resident lines of that page. 
To avoid massive tag lookups and NVM cache read operations, we introduce an alternative data layout for the NVM cache.
The proposed layout forces all the lines of a given page to be placed in a single physical row
of the LLC. A physical row contains multiple cache sets, each with multiple cache lines, and 
each physical row may contain lines from multiple pages. 

For the \arch{} LLC, we assume a physically distributed, logically shared L3 cache that acts as a victim of private L2 caches.
While we evaluate a specific design point, the design of \arch{} itself does not preclude other potential organizations of the cache hierarchy.
The cache is split into equally-sized slices. Each slice has its own controller and can independently service any type of request. 
We use STT-RAM for the data array and SRAM for the tag array for two reasons.
First, the tag array is much smaller than the data array and  its relative area overhead is small. 
Second, the LLC is highly associative, and the tags are accessed before the data array to minimize the dynamic energy of the data array access. Having NVM tags would add significant
latency to all accesses.

The LLC tag array supports both conventional accesses (e.g., triggered by L2 misses) and 4KB-aligned page-level data transfer requests triggered by L1 TLB fills. We distinguish the two by referring to the former as Cache Line Requests (CLR) and to the latter as Page Transfer Requests (PTR).

\begin{figure}[h]
  \centering
  \begin{center}
      \vspace{-3.2mm}
    \hbox{\hspace{0em}\includegraphics[width=1.0\linewidth]{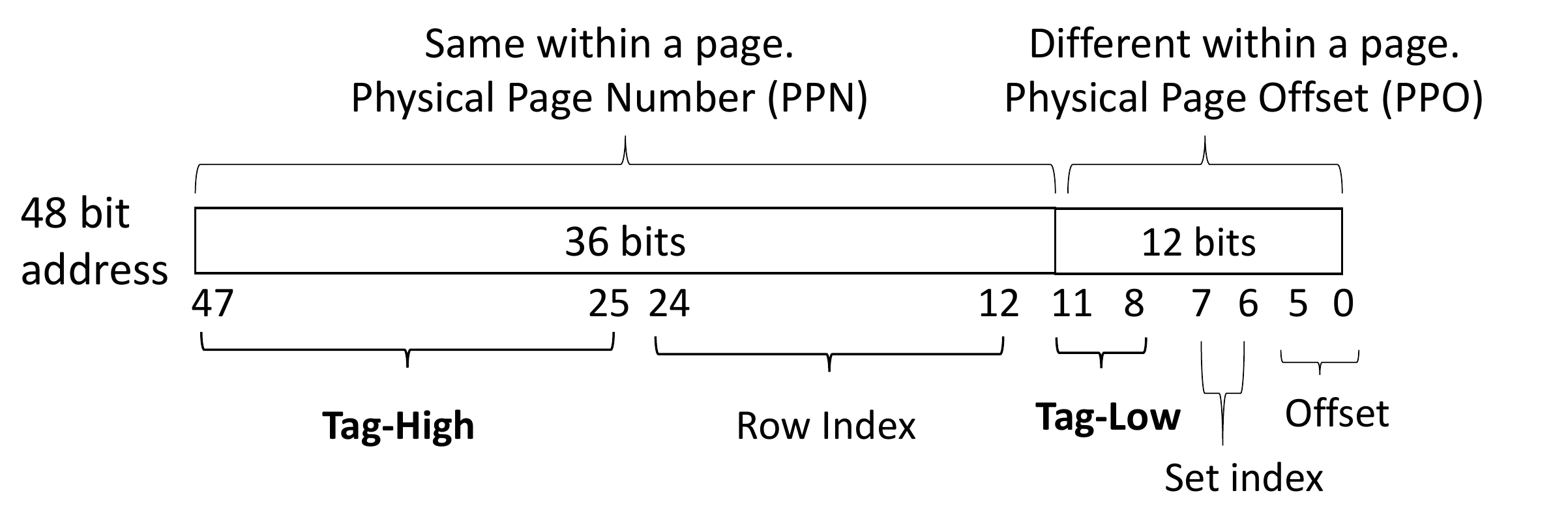}}
    \vspace{-3mm}
  \caption{\arch\ LLC addressing scheme.}
  \label{fig:addressing}
  \end{center}
  \vspace{-7.2mm}
\end{figure}

By placing all the lines of a given page in the same LLC physical row, we limit tag searching 
to only 64 entries for every incoming PTR (assuming 4KB pages and a 64B line size). 
We alter the LLC cache indexing to support our new data layout for both PTR and CLR accesses. Our addressing scheme is presented in Figure \ref{fig:addressing}.
To pick the physical row, we use some bits of the Physical Page Number (PPN) called Row Index. 
Once the  row is selected,
we use a subset of the Physical Page Offset (PPO) bits called
Set Index   to select a set within the physical row. Then, some of the
PPN bits ({\em Tag-High}) and of the 
PPO bits ({\em Tag-Low})  are used as
tag. Finally, the remaining PPO bits are used as block offset (Figure~\ref{fig:addressing}).

PTRs and CLRs differ in the tag match logic. Specifically, for tag matching,
a PTR access ignores the
page offset bits. Hence, it only uses {\em Tag-High} bits.
In contrast, a CLR access uses both {\em Tag-High} and {\em Tag-Low} bits for tag matching.

The lines of a page could be split across LLC slices, mapping to the same physical row in each slice.
However, in order to simplify the tag hit logic and NVM to PB data movement, we choose to map the entire page in the same LLC slice.
Note that our layout does not impose any restrictions on the LLC organization (e.g., line size, associativity, etc.).

\noindent
{\bf Example.} To illustrate the proposed layout, we show an example with a
single-slice of 32MB size, 16-way set-associative LLC with 64B cache lines and 4KB pages. 
The LLC has 8192 physical rows, each organized in 4 sets, 16 ways each. 
Each physical row is 4KB and can be banked if needed.
As shown in Figure~\ref{fig:addressing}, the physical address (PA)
has 48  bits, the 12 least significant ones are the PPO, and
bits 0-5 are the line offset.

The cache index bits include the row index bits (bits 24:12), which select the row of the cache, and the set index bits (bits 7:6), which select the cache set within a row.
Note that the row index bits (bits 24:12) do not include any PPO bits.
Bits 11:8 and 47:25 form the tag, split into Tag-Low and Tag-High parts, respectively.
For tag matching, a CLR selects a row and a set using indexing bits 24:12 and 7:6,
respectively, and finds a match using the tag bits (bits 47:25 and 11:8). 
For tag matching, a PTR selects a row using the row index bits (bits 24:12) and finds all matches using the subset of tag bits lying outside the PPO, namely the Tag-High (bits 47:25).
Thanks to this data layout, the PTR does not search the entire cache; it only checks 
the Tag-High (bits 47:25) of the 64 lines in the selected cache row. 

The dynamic energy of a CLR tag access is proportional to the 16 lines $\times$ 27 tag bits comparison (432 bits). The dynamic energy of a PTR tag access is proportional to the 64 lines $\times$ 23 tag bits comparison (1,472 bits). A PTR tag access consumes 3.4$\times$ more energy than a conventional CLR tag access. 
Triggering PTR tag searches only on TLB re-filled pages lowers the overall energy cost of accessing the PBs. 

Our data layout could be extended to optimize for huge pages.
However, we find that such a design is not efficient, as huge pages increase the overhead of tag lookup and data movement, while it is unlikely that a large fraction of their lines will be LLC-resident.
In Section \ref{huge_pages}, we present how we efficiently handle huge pages.

\subsection{Promotion of Cache Lines to Page Buffers}

\arch{} populates each PB with LLC-resident cache lines from the same page.
The process is as follows.
Once a PTR reaches the LLC controller, the hardware checks if any PB already has
lines from the accessed page. If not, the hardware
 checks the LLC tags to find if the LLC holds more than a threshold
 number of cache lines of the accessed page. If so, the 
 LLC-resident lines of the
page are transferred  to a PB. 
The NVM cache is inclusive of the PBs. Hence, the transferred lines are not  invalidated from the NVM data array.

\arch{} provides hardware to bring cache lines to the PBs.
According to Figure \ref{fig:l3resident}, the LLC may only contain
a few  of the lines 
from a given page. Hence, PBs will be sparsely populated.
In order to increase PB utilization, it is preferable to have more and smaller PBs.
For this reason, we propose PBs that are smaller than a page. 
However, given that the size of a physical row is equal to a page, steering the data from a row to a
PB is not trivial and requires keeping metadata. 
To simplify both the metadata and the routing overhead, we propose using PBs 
of size equal to half a page (2KB), and multiplex the two halves of a page into the same
PB.

\begin{figure}[ht]
  \centering
  \begin{center}
      \vspace{-0mm}
    \hbox{\hspace{0.4em}\includegraphics[width=0.95\linewidth]{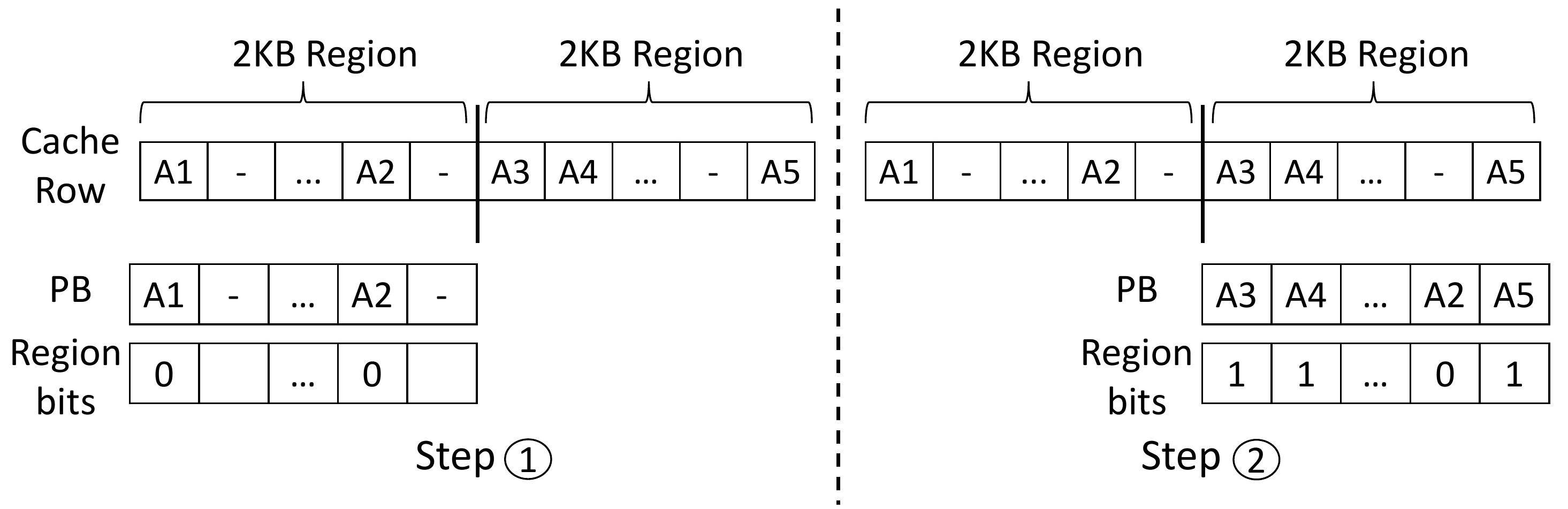}}
  \vspace{-1.9mm}
  \caption{Promotion of the lines of a page to a PB.}
  \label{fig:pb_activation_example}
  \end{center}
  \vspace{-7.9mm}
\end{figure}

\noindent
{\bf Example.}
Figure~\ref{fig:pb_activation_example} shows an example that promotes cache lines from page \textit{A} into a PB. Since we have 4KB pages (and, hence, 
4KB physical rows) and use 2KB PBs, we logically partition a physical
row into two 2KB regions, and promote the lines of a page
into the PB in two steps:
first from the leftmost region of the physical row, and then from
the rightmost region of the row. 

The left side of Figure~\ref{fig:pb_activation_example} shows the first step.
At the top, we see the the state of the cache row. The leftmost region has
lines in its first position ({\em A1}) and in the one before last ({\em A2}).
Hence, we promote these lines into the PB.
To simplify the routing, as shown in the figure, 
the lines are placed in the PB in the same slots that they use in the 2KB region. 
The PB does not need to store any address tags because any memory
access will check the SRAM LLC tags first, to decide whether the 
corresponding PB line is valid.
However, each PB slot has a bit to identify
which region the line comes from. This bit is needed
to fully identify the line. In the example, since the two lines come from 
the leftmost region, the bits are  0. In our example, we need 32 such bits, which we call {\em Region} bits.

The right side of Figure~\ref{fig:pb_activation_example} shows 
the second step. The top part repeats the cache row state. The 
rightmost region has
lines in its first ({\em A3}), second ({\em A4}), and last ({\em A5})
positions. Hence, we promote these lines into the PB and set the
Region bits of the entries to 1.

Note that the two lines in the first position of the two regions wanted to
use the same PB slot, and we had to pick a winner. In the example, 
we picked {\em A3} over {\em A1}. To pick a winner, \arch{}
uses a simple 
algorithm
that tries to guess which of the two lines is more likely
to be used in the future. Specifically, \arch{} takes the address {\em A} 
that triggered the TLB miss
and records whether {\em A} belongs to the first or second half of the page.
Then, when populating a PB, on a conflict in a PB entry, 
the line from the same half of the page as {\em A} overwrites the
line from the different half of the page. This algorithm guesses that,
because of spatial locality, the former is more likely to be accessed
soon that the latter.

Thanks to LLC's organization, the operation of 
promoting the lines from the two regions (and, in another design,
from potentially more regions) into a PB 
does not stall the LLC pipeline more than a single read. Indeed,
all the cache lines of a page are on the {\em same physical row}, 
and thus they are promoted to a PB with {\em  a single read operation} in the NVM data array.
The writes into the PB are also pipelined:
as the first region is written, the second performs the checks.

\subsection{Tag Checks}

\begin{figure}[ht]
  \centering
  \begin{center}
    \vspace{-3.5mm}
    \hbox{\hspace{1em}\includegraphics[width=0.9\linewidth]{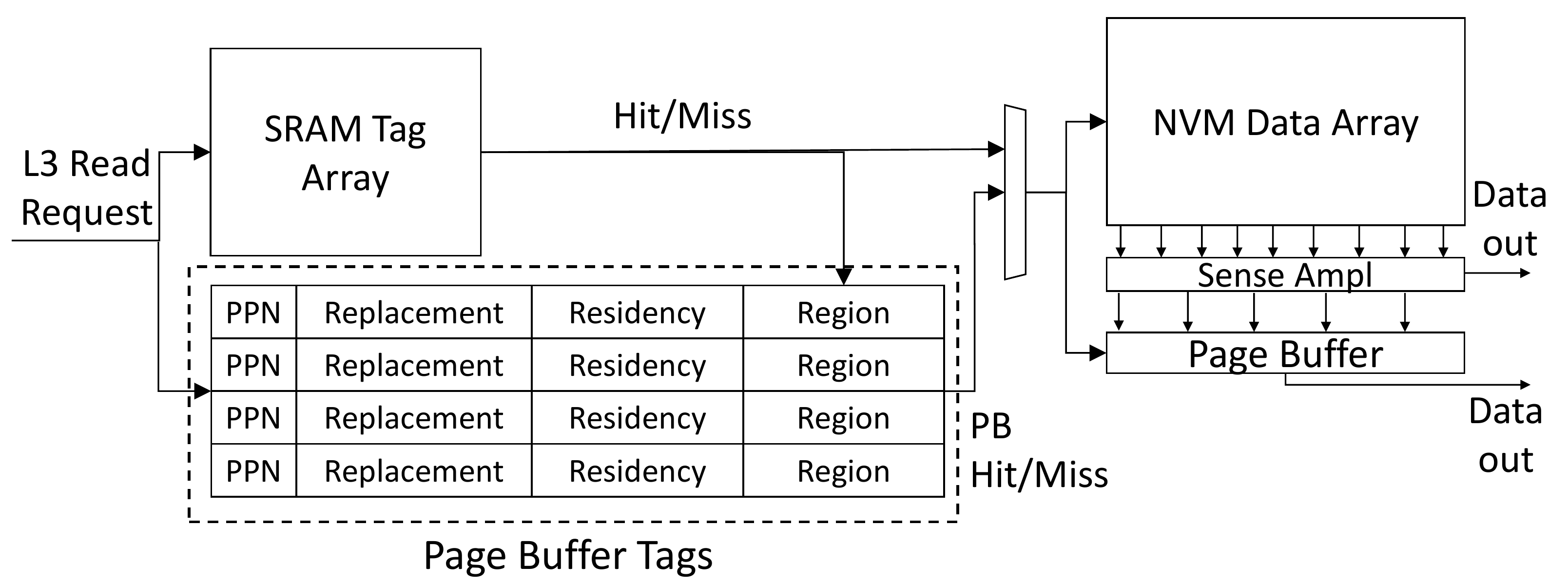}}
    \vspace{-1.2mm}
  \caption{LLC read request path in \arch{}.}
  \label{fig:l3_access}
  \end{center}
  \vspace{-6.7mm}
\end{figure}

To keep track of the pages and cache lines that are present in the PBs, 
\arch{}  has a hardware structure organized as an array called the {\em Page Buffer Tags} (PB Tags) (Figure \ref{fig:l3_access}). Each entry in the PB Tags corresponds to one PB.
An entry has: (a) the PPN of the page whose lines are stored in the PB,
(b) a {\em Replacement} counter to manage PB replacement, (c) a {\em Residency} counter that tracks the number of valid lines in the PB, and 
(d) the {\em Region} bits discussed above.

Both PTR and CLR use the
PB Tags to determine whether a PB 
contains data for the page requested. 
In the case of a PTR, the PB Tags are checked to identify if 
there is an entry with the PPN of the requested page.
In a CLR, the PB Tags are checked to identify if there is 
an entry with the
requested PPN and with the correct bit in the Region field corresponding
to the requested line.

The operation of a CLR 
is as follows. The CLR
accesses both the LLC tag array and PB Tags simultaneously (Figure~\ref{fig:l3_access}). 
It uses the PPN bits of the PB Tags to identify if a PB contains 
lines from the page accessed. 
It uses the LLC tag array to identify whether and where
the requested line resides in the LLC physical row.
If the address of the line is not found in the LLC tag array, an LLC miss is declared.

However, if the LLC tag array indicates a cache hit, \arch{} checks for a PB hit. A PB hit will occur if the region of the physical row with
the matching address is the same as the one indicated by
the Region bit of the corresponding location of the PB. In this case,
the line is accessed from the PB in the same position.  Recall that, during data transfer,
lines were moved from the LLC to the PB without reordering. 
If the Region bit does not match or the PPN bits do not match,
the line is accessed from the NVM-LLC data array. 

Note that the access to the PPNs in the PB tags overlaps with the 
access to the LLC tag array. The access to the Region bit in the 
PB tags is only performed after the LLC tag array access
(Figure \ref{fig:l3_access}). However, 
accessing the Region bit
only extends the critical path by  one cycle (when 
both the PPN and the LLC tag array hit).

The PB contents are always kept synchronized with the NVM LLC. 
When a line is written to the LLC, the corresponding slot in a
PB,
if present, is updated.
For this reason, there is no need to write back PBs to the 
NVM cache. There is also no need to keep valid or coherence state bits in PB Tags because the LLC tags provide such information.
Whenever an LLC line is invalidated (due to an external probe or L2 promotion),
or evicted (due to an LLC replacement), the corresponding valid bit of the line 
in the LLC tags is reset. No other action is needed: given that the PB hit logic waits for the 
LLC tag search to complete, a CLR will not read the PB slot data if its corresponding 
LLC line is invalid, even if the data is still resident in the PB entry. 

The Residency counter of a PB in the PB Tags tells how many cache lines are valid in the PB. 
This counter is set when the lines of a page are moved from the NVM
cache to the PB.
It is decremented when one of its lines is invalidated or evicted
from the LLC. 
It is incremented when an L2 victim is installed in the LLC and copied
into the PB.
The Residency and Replacement counters are used to handle PB replacement,
as we discuss in 
Section~\ref{replacement_policy}.

\noindent
{\bf Example.}
The PBs add little area overhead to the LLC.
To see why, consider an example based on Figure~\ref{fig:addressing}. 
A PB is composed of  tag and data.
For the tag, we have a 36-bit PPN and assume a 
10-bit Replacement counter. The Residency counter needs 
\(log_{2}(PB size / line size)\) bits, which is 5 in our example.
The Region bits are 
\(PB size / line size * log_{2}(4KB/PB size)\), which is 32 in our case. The total comes to 83 bits per PB tag entry. 
We then add  the
size of the PB data, which is 2KB. Based on this data,  
each PB adds a $\approx$0.05\% area overhead over a 16MB LLC NVM slice.
If we 
assume a 4:1 area ratio between NVM and SRAM, the
area overhead over the total LLC slice (data array and tags)
is $\approx$0.046\%.

\subsection{Page Buffer  Replacement Policy}\label{replacement_policy}
Sometimes, \arch{} needs to find an available PB to promote a page's cache lines, and all PBs are in use.
To pick a PB, \arch{} uses a PB replacement algorithm that
considers: (i) how many cache lines are resident in a PB, and (ii) the frequency of accesses to the page in the PB. 
The goal is to capture the dynamic behavior of accesses to  
each PB, and neither replace a PB too early (before 
its entries are accessed), nor keep a page resident in the PB if it 
is not being accessed.

Specifically, when a PB is loaded, its Replacement counter is set to
the product of the Residency counter and a programmable constant called {\em Activation Period}. 
At every cycle, the Replacement counter is decreased by one. When the PB
is accessed, either for a read or a write operation, 
the Replacement counter is recalculated by multiplying the 
current value of the Residency counter with the Activation Period. 
Furthermore, PB accesses change the Residency counter. 
On a PB read, \arch{} decrements
the Residency counter because a line from the PB is moved to
the private caches. 
On a PB write, \arch{} increments the Residency counter 
because a line is written back
from L2. Once the Replacement counter reaches zero, this PB is subject to replacement.

\subsection{Huge Page Management} \label{huge_pages}
\vspace{-0.2mm}
Modern systems support huge pages, such as 2MB and 1GB, to alleviate TLB pressure.
Even though we described \arch{} in the context of 
4KB pages, 
\arch{} can support huge pages without any modification to the NVM cache layout.

We envision a physical row in the LLC cache to still hold 4KB of data. 
However, if we chose to transfer a whole 2MB page into a PB, we would need the costly search of many rows. Moreover, a large-sized PB would likely be underutilized.

Consequently, we use a different design where \arch{} only transfers individual
4KB chunks of data at a time from a huge page into a PB. Specifically, when 
a huge page entry is re-filled into the L1 TLB, \arch{} only brings 
into a PB the 4KB chunk
of this huge page that contains the address that triggered the TLB miss. 
In addition, the L1 TLB records this 4KB chunk that triggered the TLB miss.
Subsequently, when an access to the same huge page, but a different 4KB chunk occurs,
\arch{} triggers a transfer of the new 4KB chunk, and again records the chunk in the TLB. In this way, \arch{} can have
multiple 4KB chunks of the same huge page active in the PBs.

\arch{} adds this support for  2MB and 1GB pages. 
For the 2MB pages, \arch{} needs to add 9 bits per L1 TLB entry to 
record the most-recently-promoted 4KB chunk. For the 1GB pages,
\arch{} needs to add 18 bits per L1 TLB entry. 
These are  minimal overheads. 

\vspace{-0mm}
\section{Evaluation Methodology}
\vspace{-0.5mm}
\subsection{Modeled Architecture and Infrastructure} \label{ModelDetails}

\begin{table}[ht!]
\vspace{-0.mm}
\caption{Architectural parameters used for evaluation. In the 
table, RT means round trip latency from the core.}
\vspace{-1mm}
\label{table:architecture}
\centering
\resizebox{1.0\linewidth}{!}{
\begin{tabular}{||l|l||}
\hline\hline
\multicolumn{2}{||c||}{Processor Parameters}\\
\hline\hline
Multicore chip  &  4 OoO cores, 4-issue, 22nm, 3.2GHz\\
Ld-St queue; ROB & 92 entries; 192 entries\\
L1 cache  &  32KB, 8-way, 2 cycles round trip latency (RT), 64B line\\
L2 cache  &  512KB, 8-way, 14 cycles RT, 64B line\\
Prefetchers & L1 cache: stride prefetch.ß L2 cache: next-block prefetch \\
LLC SRAM cache &  4MB/core, 16-way, 1 slice/core, 64B line\\
         &   53 cycles RT, 2 cycles tag latency, 12 cycles data latency \\
         &   Energy: Read/Write 0.47/0.48nJ, Tags 4pJ, Leak: 1.4W \\
L1 TLB &  64 entries, 4-way, 2 cycles RT\\
L2 TLB &  1024 entries, 12-way, 12 cycles RT \\
\hline\hline
\multicolumn{2}{||c||}{NVM cache parameters}    \\
\hline\hline
LLC NVM cache & 16MB/core, 16-way, 1 slice/core, 64B line\\
         &   63 cycles RT read latency, 78 cycles RT write latency \\
         &   2 cycles tag access latency, 22 cycles data access latency  \\
         &  (of which 10 cycles are not  pipelined) \\
         & Energy: Read/Write 0.95/6.3nJ, Tag 7pJ , Leak: 829mW \\
Page Buffers (PB) & 20 PBs, 2KB/each, 43 cycles RT \\
                  & Energy: Read/Write 12/13pJ, Tag 12pJ, Leak: 4.1mW  \\
PTR signal latency & 6 cycles\\
NVM cache to PB threshold & 6 cache lines\\
PB activation period      & 20 cycles per active cache line\\
PB area overhead & 1\% area overhead over LLC data array (0.05\% per PB)\\
\hline\hline
\multicolumn{2}{||c||}{Main-Memory Parameters}    \\
\hline\hline
Capacity & 64GB\\ 
Channels; Banks & 2; 8\\
Latency & 190 cycles RT (on average)\\
Freq; Bus width & 1.6GHz DDR; 64 bits per channel \\
\hline\hline
\multicolumn{2}{||c||}{System Parameters}    \\
\hline\hline
Host OS& Ubuntu Server 16.04\\
\hline\hline
\end{tabular}}
\vspace{-2.mm}
\end{table}

We use full-system cycle-level simulations to model a server architecture with
4 cores and 64 GB of main memory. 
The main architecture parameters
are shown in Table~\ref{table:architecture}. 
Each core is out-of-order with private L1 and L2 caches, and a shared 
LLC. The L1 and L2 caches use the stride and next-block prefetchers, respectively, as implemented by the SST \cite{sst} framework.
The L2 cache is inclusive of L1, while the LLC  is populated by L2 victims.
The baseline system uses an SRAM-based physically distributed, logically shared, victim LLC. For \arch{},
we modeled an increased-latency, non-pipelined NVM LLC and the 
required hardware modifications. 
We use the published data (Table \ref{table:latencies}) to estimate the minimum read latency of the 
NVM LLC data array (i.e., 3 ns).
Note that the  higher NVM latency as presented in Table \ref{table:latencies} refers to the NVM LLC {\em data array} (i.e., 10  cycles at 3.2GHz), and not the {\em total round trip latency} to access
the LLC 
from the core (as shown
in Table~\ref{table:architecture}).
There are  private L1 and L2 data TLBs, and a page walker per core.
For our evaluation, we
integrate the Simics full-system simulator~\cite{simics} with the SST~\cite{sst} framework.
To model 64GB of main memory, we used the DRAMSim2~\cite{dramsim} memory simulator.
We use Intel SAE \cite{sae} on top of Simics for OS instrumentation. 
Finally, we use CACTI \cite{cacti} and McPAT \cite{mcpat} to calculate the timing and energy parameters of our processor, all SRAM-based tag and data arrays required by \arch{}, and the Baseline SRAM-based LLC. We scaled the NVM cache energy parameters according to prior work \cite{adaptive_placement, oap}.
We model one extra clock for determining a PB hit/miss because it needs the information from the LLC tag search to determine the location of a line inside a PB.

\vspace{-0.5mm}
\subsection{Configurations and Workloads}

We compare four different design configurations.

\noindent
{\em \textbf{Baseline:} } SRAM-based LLC with the latency and size parameters described in Table \ref{table:architecture}.

\noindent
{\em \textbf{NVM-Only:} } LLC with STT-RAM for the data array
and SRAM for the tag array with the parameters of Table \ref{table:architecture}, but without PB support and with conventional indexing.

\noindent
{\em \textbf{\arch{}:} } LLC with STT-RAM for the data array and SRAM for the tag array with the proposed data layout and PB support.

\noindent
{\em \textbf{O-SRAM\footnote{This is not to be confused with Oblivious RAM (ORAM).}:} }
Optimistic hybrid design with conventional indexing, pipelined access latency and energy characteristics of \textit{Baseline}, combined with STT-RAM area density.

To evaluate the efficacy of our design, we execute 14 different benchmarks.
The benchmarks are shown in Table \ref{table:workloads} with their memory footprint and the L2 misses per kilo instructions (MPKI).
We chose ten benchmarks from the SPEC CPU\textsuperscript{\textregistered} 2017 (Group A) \cite{spec17} and SPEC CPU\textsuperscript{\textregistered} 2006 (Group B) \cite{spec}\footnote{SPEC\textsuperscript{\textregistered} and SPEC CPU\textsuperscript{\textregistered} are registered trademarks of the Standard Performance Evaluation Corporation. See www.spec.org for more information.}
benchmark suites with high MPKI that can stress the memory subsystem.
We also run four benchmarks from the CORAL \cite{coral} and CORAL2 \cite{coral2} suites (Group C), which are representative benchmarks of HPC systems. 
The memory footprint of our benchmarks does not fit entirely into the private caches and can stress the LLC. 
We select the region of interest (ROI) with SimPoint \cite{simpoint} for the SPEC\textsuperscript{\textregistered} workloads and we instrument the source code for the others. Starting from a checkpoint inside the ROI,
we warm-up the architectural state by running 500 million instructions before simulating 1.5 billion instructions.

\begin{table}[h!]
\footnotesize
\centering
\scriptsize
\vspace{-2.5mm}
\caption{Workloads.}
\vspace{-2.3mm}
\label{table:workloads}
\fontsize{7}{7.8}\selectfont 
\begin{tabular}{||c c c||c c c||}
\hline\hline
Workload & Footprint & L2   &  Workload & Footprint & L2 \\
         & (MB)      & MPKI &            & (MB)      & MPKI     \\
 \hline\hline
\multicolumn{3}{||c||}{\underline{\textbf{Group A}}} & \multicolumn{3}{c||}{\underline{\textbf{Group B}}}\\
505.mcf\_r       &   613  &  39   & & & \\
519.lbm\_r       &   409  &  10   &  450.soplex      &   436  &  10 \\
557.xz\_r        &   800  &   3   &  459.GemsFDTD    &   146  &  4 \\
\cline{1-3}\multicolumn{3}{||c||}{\underline{\textbf{Group C}}} &  473.astar       &   372  &  18  \\
Kripke & 608  &  39             & 462.libquantum  &   267  &  11 \\
XSBench         & 110  &  63    &  433.milc        &   123  &  7   \\
QLA              & 375  &  11   &  471.omnetpp     &   388  &  12  \\
lulesh           & 110  &  15   &  437.leslie3d    &   62   &  3\\ 

\hline\hline
\end{tabular}

\vspace{-4mm}
\end{table}
\vspace{-1.mm}
\section{Evaluation}

\vspace{-0mm}
\subsection{\arch{} Performance and Energy}

\begin{figure*}[ht!]
    \captionsetup[subfloat]{captionskip=0mm}
    \centering
    \subfloat[Normalized L2 miss response time - w/ 4KB Pages]{\label{fig:perf_b}\includegraphics[width=1.015\columnwidth]{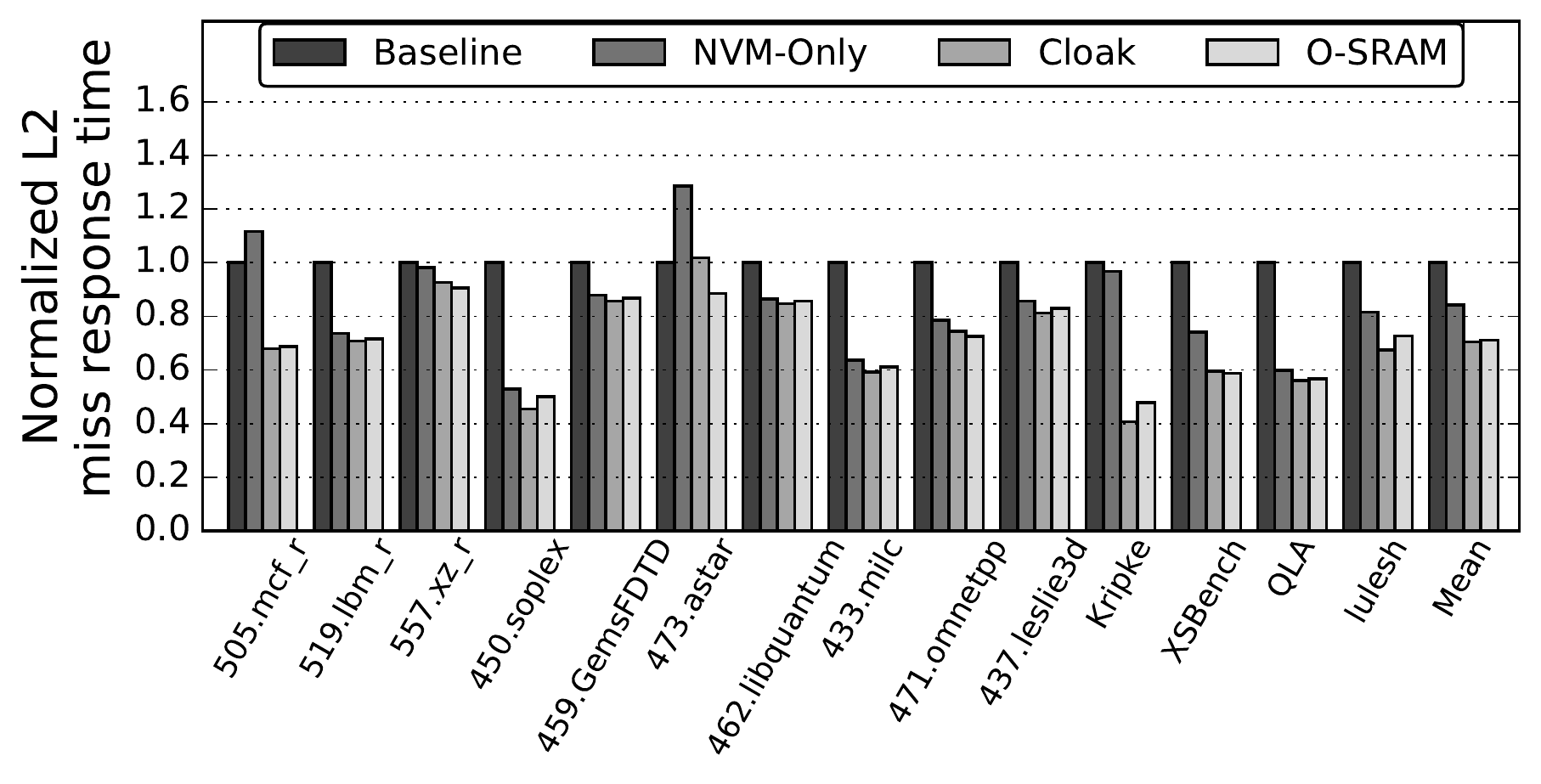}}
    \quad
    \vspace{-1.5mm}
    \subfloat[Speedup - w/ 4KB Pages]{\label{fig:perf_a}\includegraphics[width=0.985\columnwidth]{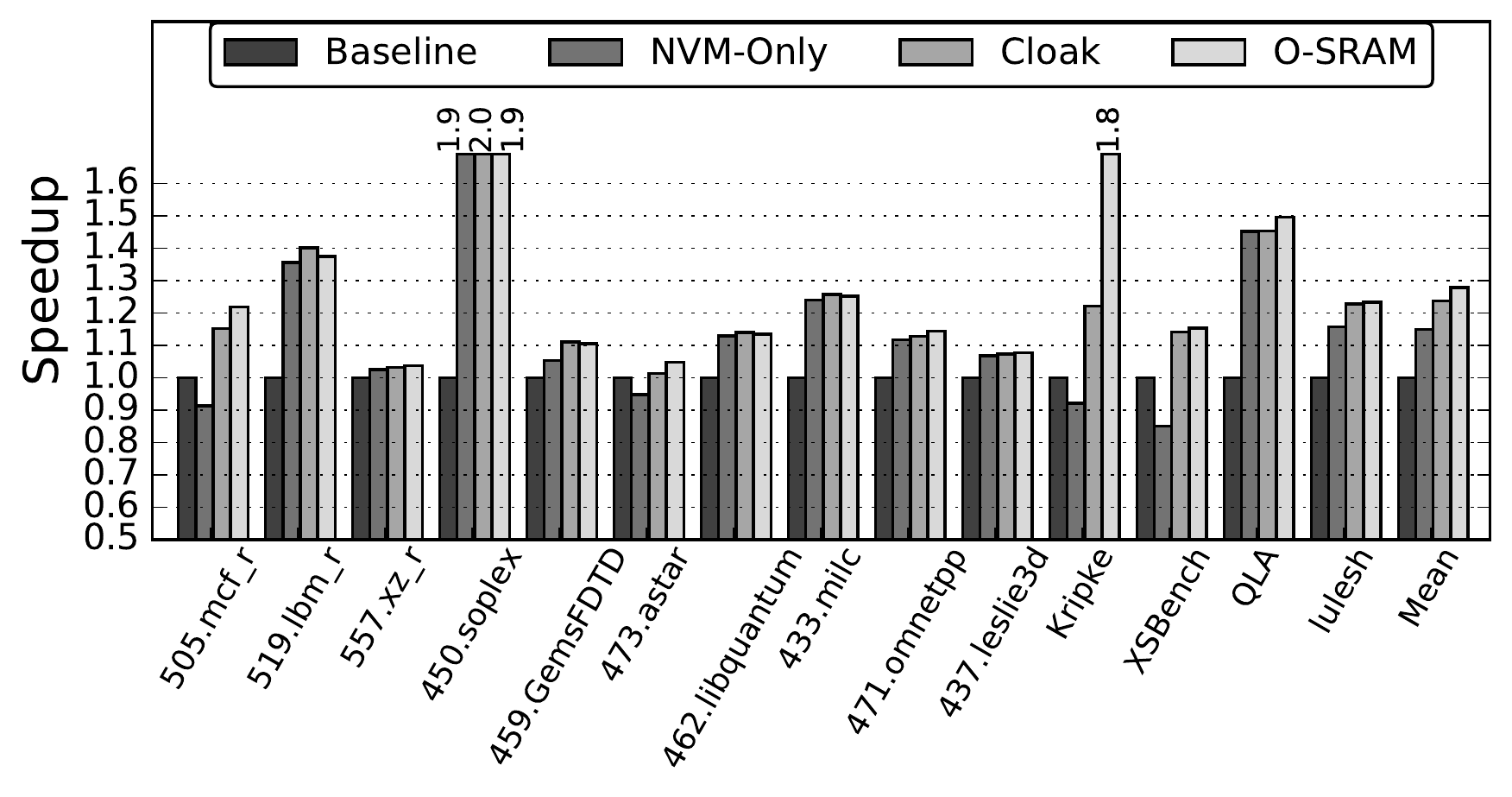}}
    \vspace{-2.2mm}
    \linebreak
    \subfloat[Normalized L2 miss response time - w/ 4KB+Huge Pages]{\label{fig:perf_d}\includegraphics[width=1.015\columnwidth]{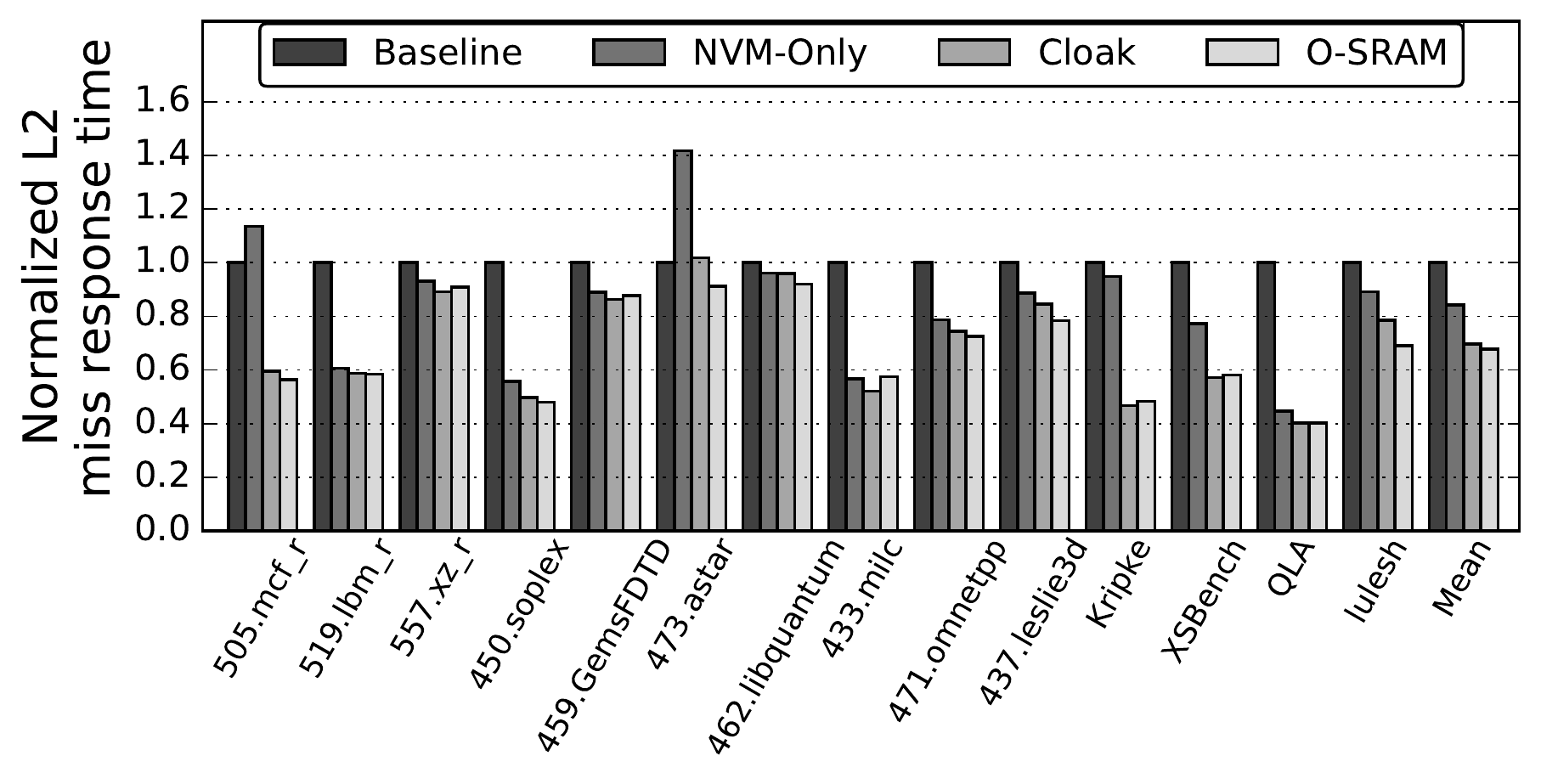}}
    \quad
    \subfloat[Speedup - w/ 4KB+Huge Pages]{\label{fig:perf_c}\includegraphics[width=0.985\columnwidth]{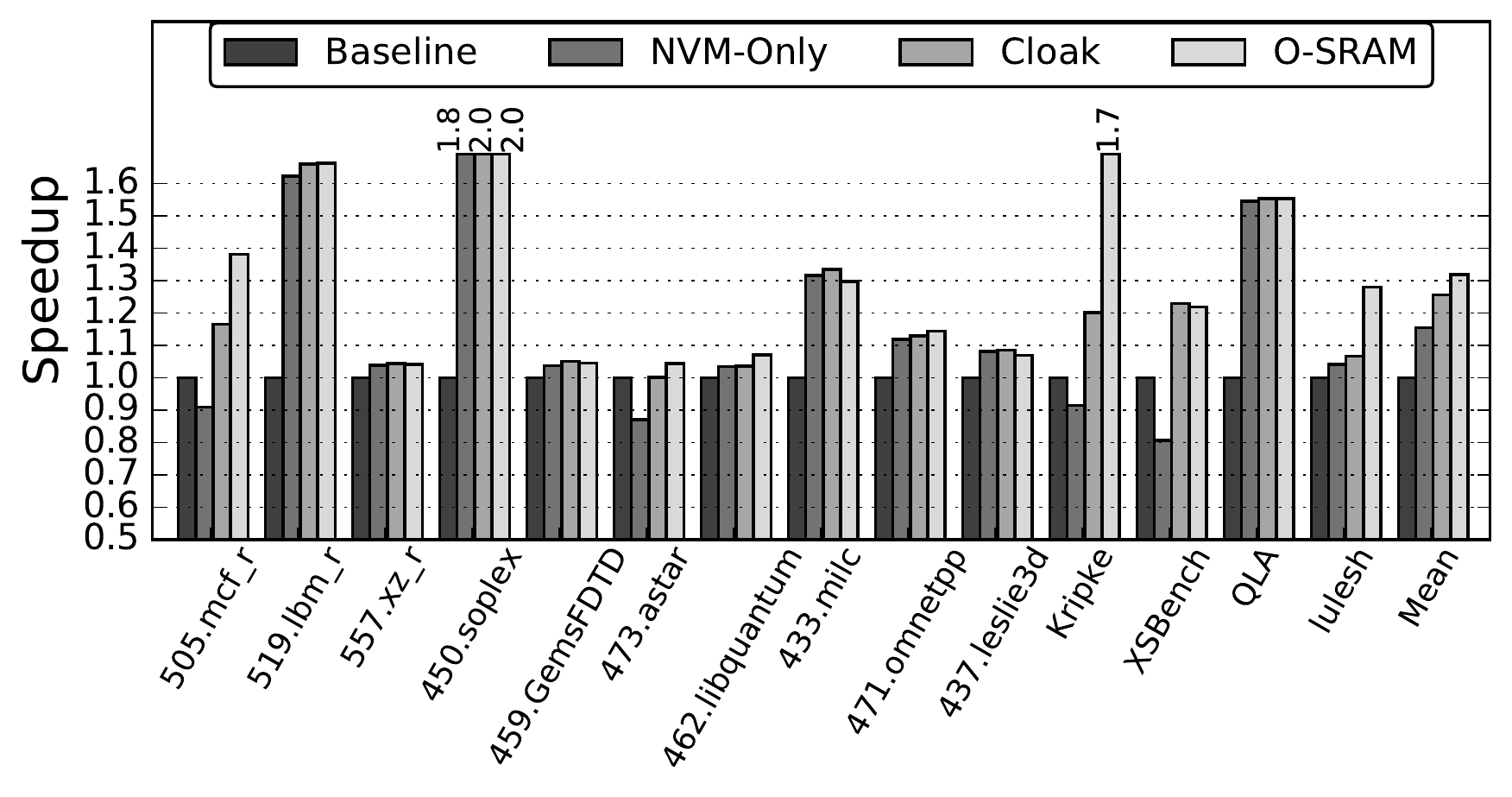}}
    \vspace{-1mm}
    \caption{Performance impact of \arch\ normalized to the Baseline configuration, when huge pages are disabled (a,b) and enabled (c,d). For each case, we show the sum of the L2 miss response times (a,c) and the application speedup (b,d).}
    \label{fig:perf}
    \vspace{-4.5mm}
\end{figure*}

In this section, we evaluate the performance of \arch{}.
When replacing an SRAM-based cache with NVM, there are two factors that affect application performance.
The first is the higher read and write latencies of NVM. 
The second is the lower cache miss rate due to the higher area density of NVM technology.
We consider two different metrics in Figure \ref{fig:perf} to show the performance impact of \arch{}.
Figures \ref{fig:perf_b} and \ref{fig:perf_d} show the 
sum of L2 miss response times for 
read CLRs, while Figures \ref{fig:perf_a} and \ref{fig:perf_c} depict the application speedup over the Baseline configuration.
All  figures are normalized to the Baseline configuration.
We conducted experiments with a system that only utilizes 4KB pages, and 
with a system with Transparent Huge Pages enabled (2MB and 1GB pages).

Figures \ref{fig:perf_b} and \ref{fig:perf_d} show 
the sum of L2 miss response times for read CLRs,
 which we will call L2 miss response time. This time is calculated as the total number of cycles from issuing an L2 miss until the miss response reaches back to the L2. On average,
\arch{} reduces the L2 miss response time by 30.0\% and 30.5\% over Baseline, with and without Huge Pages. This impact is  really close to that of the O-SRAM configuration.
The NVM-Only configuration lowers the L2 miss response time by only 15.8\% and 15.9\%. 
It does not achieve the same reduction as \arch{} or O-SRAM because of its higher and non-pipelined 
LLC hit latency.
These results indicate that the PBs are effective at reducing the NVM cache read latency---practically as much as O-SRAM.

Figures \ref{fig:perf_a} and \ref{fig:perf_c} show the
application speedup over Baseline. We see that NVM-Only LLCs can increase performance.
The reason is the larger LLC capacity achieved via NVM 
technology, which can greatly decrease the LLC miss rate.
However, there are benchmarks where  NVM-Only  experiences performance degradation compared to Baseline (505.mcf\_r, 473.astar, Kripke, and XSBench), because the lower LLC miss rate cannot compensate for the higher LLC hit latency. Benchmarks with high L2 MPKI and high LLC hit rate suffer more from the increased read latency of an NVM-based LLC.
For instance, 473.astar and XSBench with Huge Pages experience 13\% and 19\% lower performance than Baseline, respectively.

On the other hand, \arch{} consistently attains higher performance than Baseline and NVM-Only. 
There are times when it even outperforms O-SRAM. 
This can happen for benchmarks with high PB hit rate because a PB hit has lower access latency than an SRAM-based LLC hit. This is due to the lower routing latency observed 
retrieving data from the PB data array compared to from a much larger SRAM-based LLC slice.
On average, \arch{} is 25.6\% and 23.8\% faster than  Baseline with and without Huge Pages, respectively, while NVM-Only is 15.5\% and 14.9\% faster than  Baseline. 
For some benchmarks, \arch{} outperforms Baseline by up to 97\%, effectively hiding the increased read latency of NVM.

We also tested \arch{}'s efficacy by running mixes of four benchmarks. It can be shown that  NVM-Only and \arch{} outperform baseline by 24\% and 31\%, without Huge pages and by 27\% and 33\% with Huge pages, respectively. 
The rest of our evaluation focuses on a system that utilizes only 4KB pages due to space limitations. However, the performance trends remain the same when huge pages are enabled.

\begin{figure}[h]
  \centering
  \begin{center}
    \vspace{-1.2mm}
    \hbox{\hspace{0em}\includegraphics[width=1\linewidth]{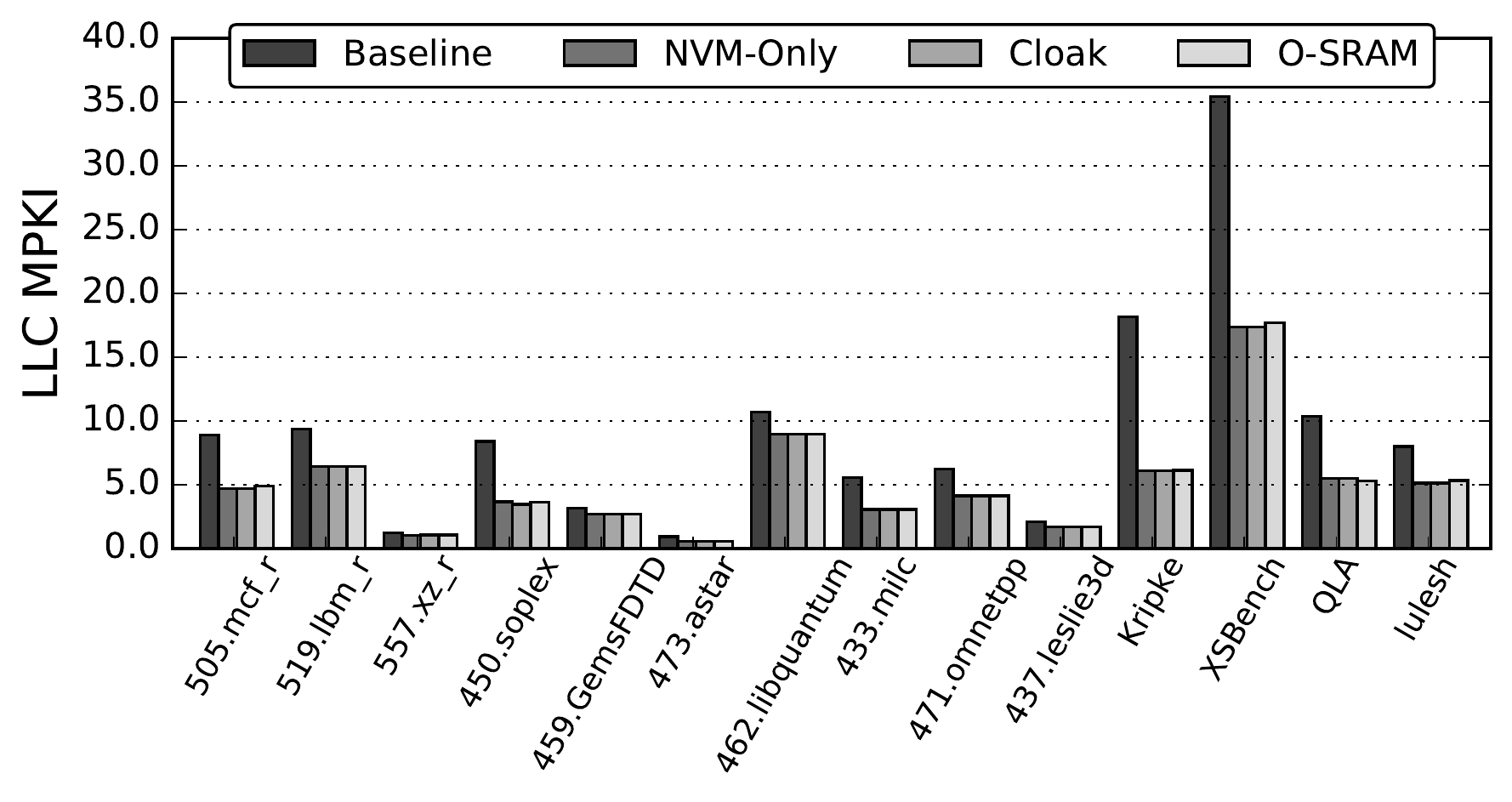}}
    \vspace{-3mm}
  \caption{LLC Misses per 
Kilo-Instructions (MPKI).}
  \label{fig:miss_rate}
  \end{center}
  \vspace{-8.2mm}
\end{figure}

\begin{figure}[h]
  \centering
  \begin{center}
    \vspace{-0mm}
    \hbox{\hspace{0em}\includegraphics[width=1\linewidth]{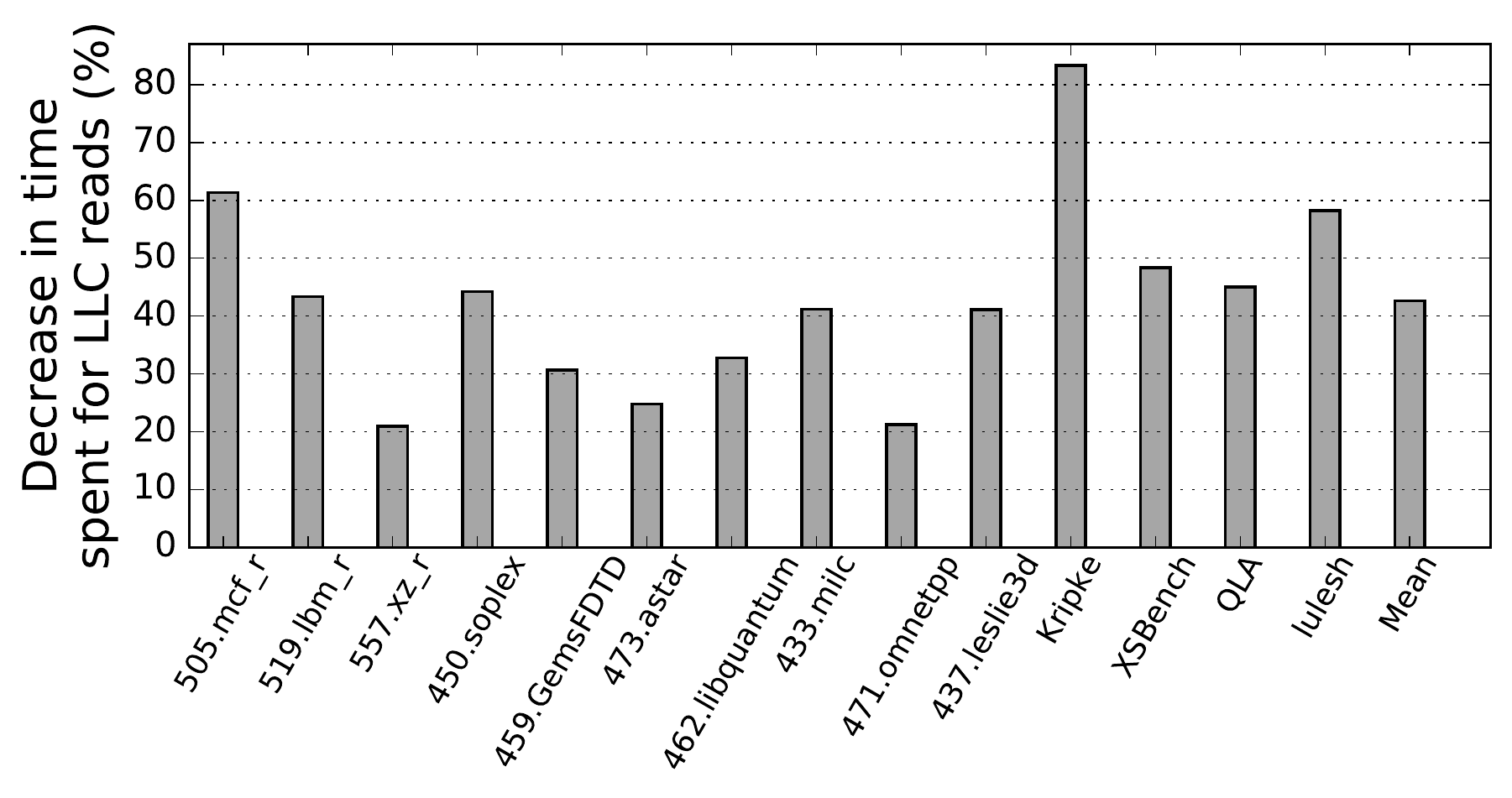}}
    \vspace{-3.1mm}
  \caption{LLC read latency reduction of \arch{} over NVM-Only.}
  \label{fig:l3_time}
  \end{center}
  \vspace{-11.4mm}
\end{figure}

To further understand the performance impact of \arch{}, we present two more performance metrics in Figures \ref{fig:miss_rate} and \ref{fig:l3_time}. In Figure \ref{fig:miss_rate}, we show the drop in LLC Misses per 
Kilo-Instructions (MPKI) for the four configurations. 
The increased LLC capacity with NVM greatly reduces
the MPKI of the applications. 
The MPKI  decreases up to 55\%, and drops across all 
benchmarks---including those with the highest MPKI
such as XSBench (50\% drop).
In most cases, \arch{} achieves an LLC MPKI close to that
of O-SRAM. 

To isolate the performance impact of PBs, Figure \ref{fig:l3_time} compares the total time that read requests spent in the LLC in NVM-Only and \arch{}.
This time is calculated as the total number of cycles from 
when an L2 miss is issued until the response is sent back to L2 from the LLC (in case of an LLC hit), or until the LLC declares
a miss (in case of an LLC miss).
Note that the two configurations have similar LLC MPKIs.
Therefore,  their cycle count difference depends on the PB hit rate in \arch{}. 
Figure \ref{fig:l3_time} shows that \arch{} notably 
reduces these LLC read latency cycles and, therefore,
accelerates the LLC read traffic.
On average, LLC CLRs spent 42.5\% less time in the
LLC with \arch{}  than with NVM-Only.
The PBs are able to speed-up \arch{}  because they service CLRs much faster than the LLC NVM-based data array.
Moreover, when CLRs are serviced from the PBs, they do not block the LLC data array pipeline, giving the opportunity to subsequent CLRs  that do not target PB-resident regions, to proceed in parallel.
As a result, the PBs not only service requests faster but they also increase the overall throughput of the LLC cache.

\begin{figure}[h]
  \centering
  \begin{center}
    \vspace{-2mm}
    \hbox{\hspace{0em}\includegraphics[width=1\linewidth]{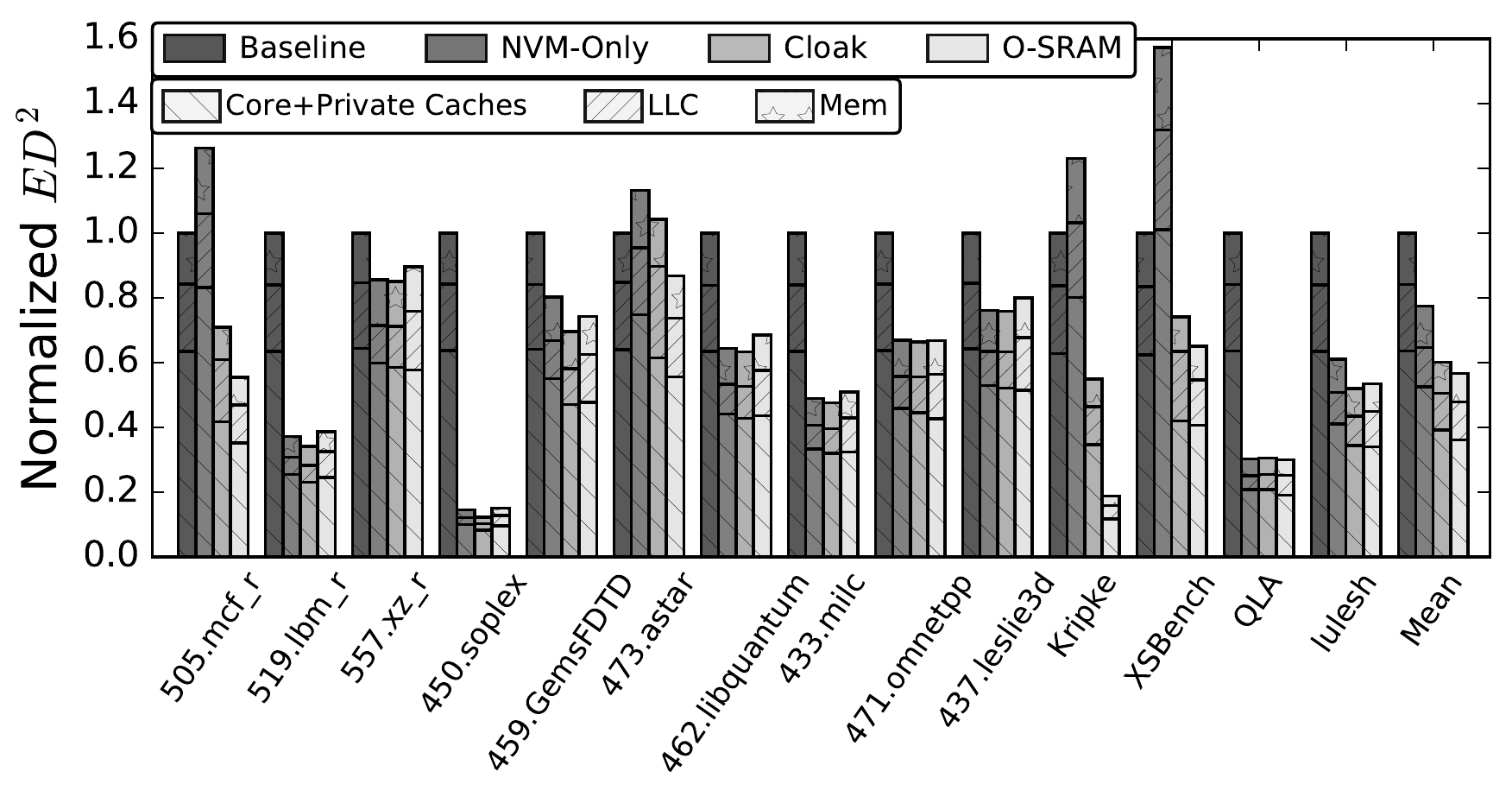}}
    \vspace{-2.6mm}
  \caption{\(ED^2\) normalized to Baseline.}
  \label{fig:energy}
  \end{center}
  \vspace{-8.5mm}
\end{figure}

Figure~\ref{fig:energy} shows the \(ED^2\) 
of the different configurations normalized to Baseline.
The bars are broken down into the contributions of the core plus private caches, the LLC, and main memory.
Overall, we see that  all the NVM designs have a lower
\(ED^2\) than Baseline. On average,
NVM-Only reduces the \(ED^2\) by 22.4\%, while \arch{} reduces it by 39.9\%. The reasons are the lower execution times of the NVM configurations, the lower leakage power of the NVM cache (which is the main energy contributor of the LLC), and the reduced number of accesses to main memory.
Compared to the NVM-Only design, it can be shown
that \arch{} consumes more energy 
in the LLC. This is because \arch{} performs more tag checks and read accesses to the NVM data array to fetch data to the PBs, a \arch{} request needs to check both the LLC and PB tags, and the PBs
consume extra leakage power.
However, this extra energy is compensated by a faster execution of \arch{} because it services many requests from the PBs.
O-SRAM reduces the \(ED^2\) by 43.3\% over Baseline on average, delivering the best efficiency. However, we see that, in some cases, \arch{} is better. O-SRAM has the same leakage power as baseline.
So, in cases where the performance of O-SRAM cannot make up for its extra leakage, \arch{} is better.

\vspace{-1.2mm}
\subsection{\arch{} Characterization}
\vspace{-0.1mm}

To achieve high performance, it is crucial to 
maximize the use of PBs.
In this section, we measure  that 84.4\% of the PTRs sent by \arch{}
to the 
LLC are for pages that have at least 6 cache lines in the LLC. 
Moreover, in 99\% of the PTRs, we are able to find a PB to 
promote the 
lines.

\begin{figure}[h]
  \centering
  \begin{center}
      \vspace{-0.0mm}
    \hbox{\hspace{0em}\includegraphics[width=1\linewidth]{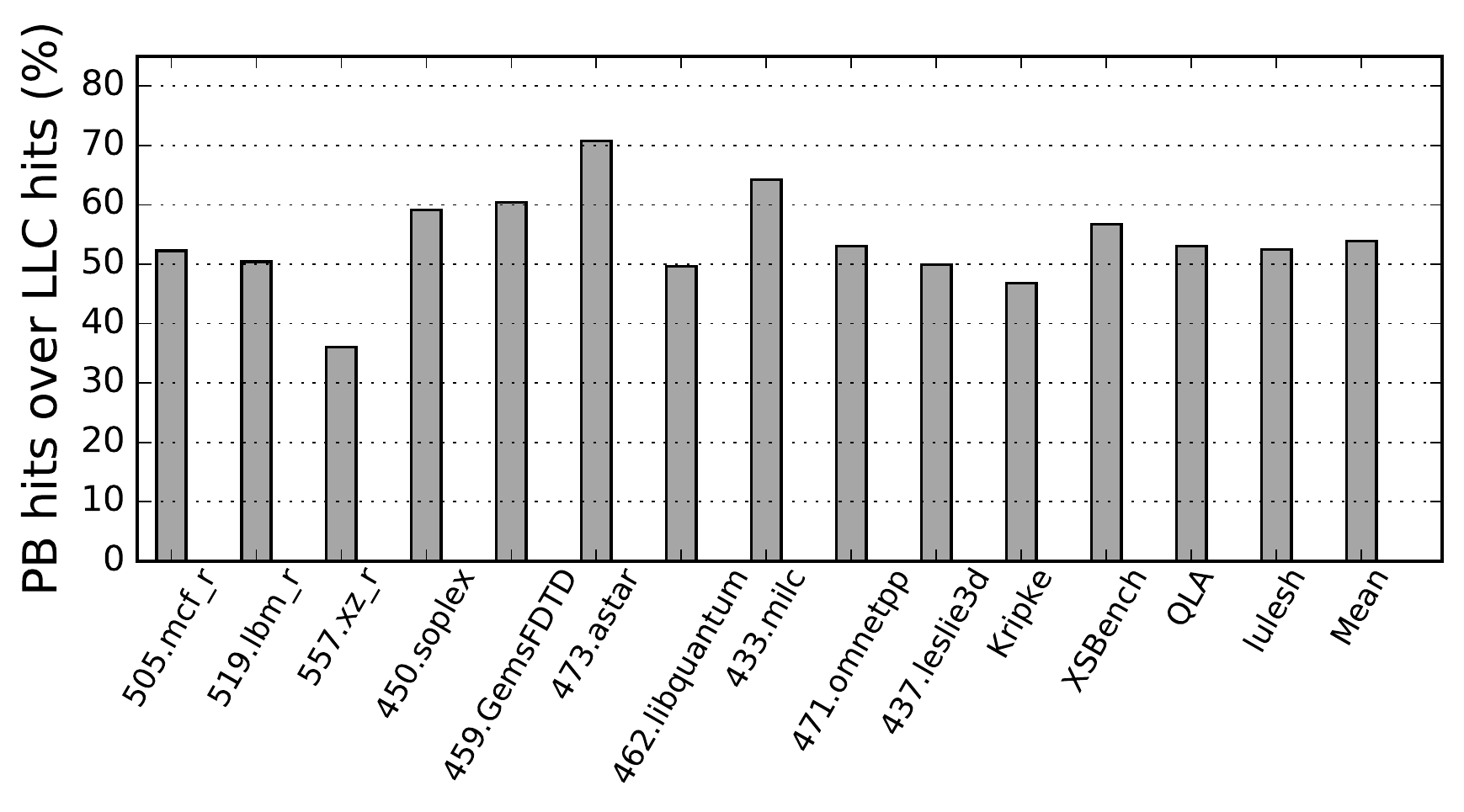}}
    \vspace{-3mm}
  \caption{Percentage of LLC hits that are serviced by PBs.}
  \label{fig:pbhits}
  \end{center}
  \vspace{-9.mm}
\end{figure}

To get further insight,
Figure~\ref{fig:pbhits} shows the percentage of LLC hits serviced 
from the PBs (instead of from the NVM data array).
On average, 54\% of the hit CLRs are serviced from the PBs
instead of from the NVM data array. 
The benchmarks with the highest LLC hit traffic, such as XSBench (45 HPKI or Hits per Kilo-Instructions) and Kripke (32 HPKI) hit in the PB 57\% and 48\% of the time, respectively. This leads to substantial performance gains of \arch{} over NVM-Only, as shown in Figure \ref{fig:perf_d}.

\begin{figure}[ht]
    \captionsetup[subfloat]{captionskip=-0mm}
    \centering
    \vspace{-4.8mm}
    \subfloat[]{\label{fig:pbhits_act}\includegraphics[width=0.505\columnwidth]{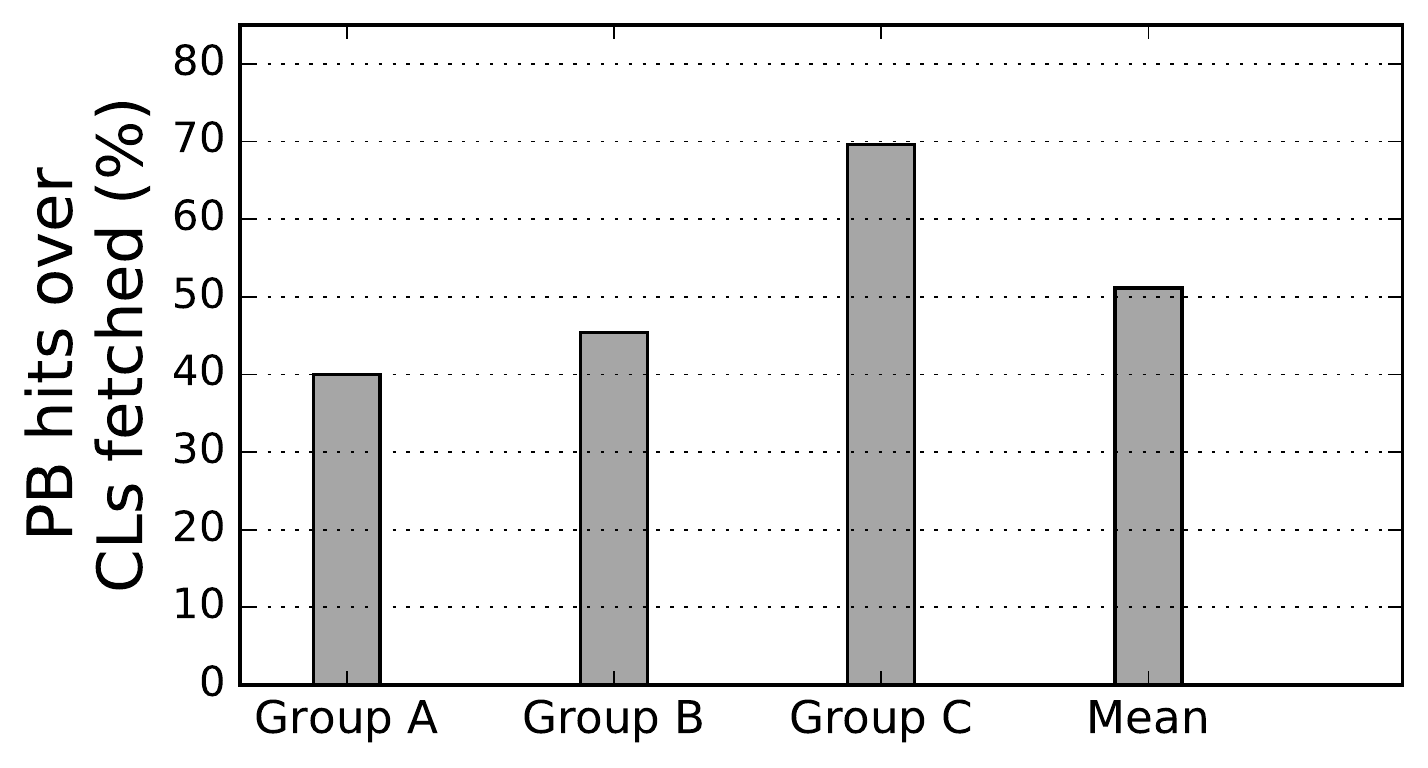}}
    \subfloat[]{\label{fig:acts_per_hint}\includegraphics[width=0.5\columnwidth]{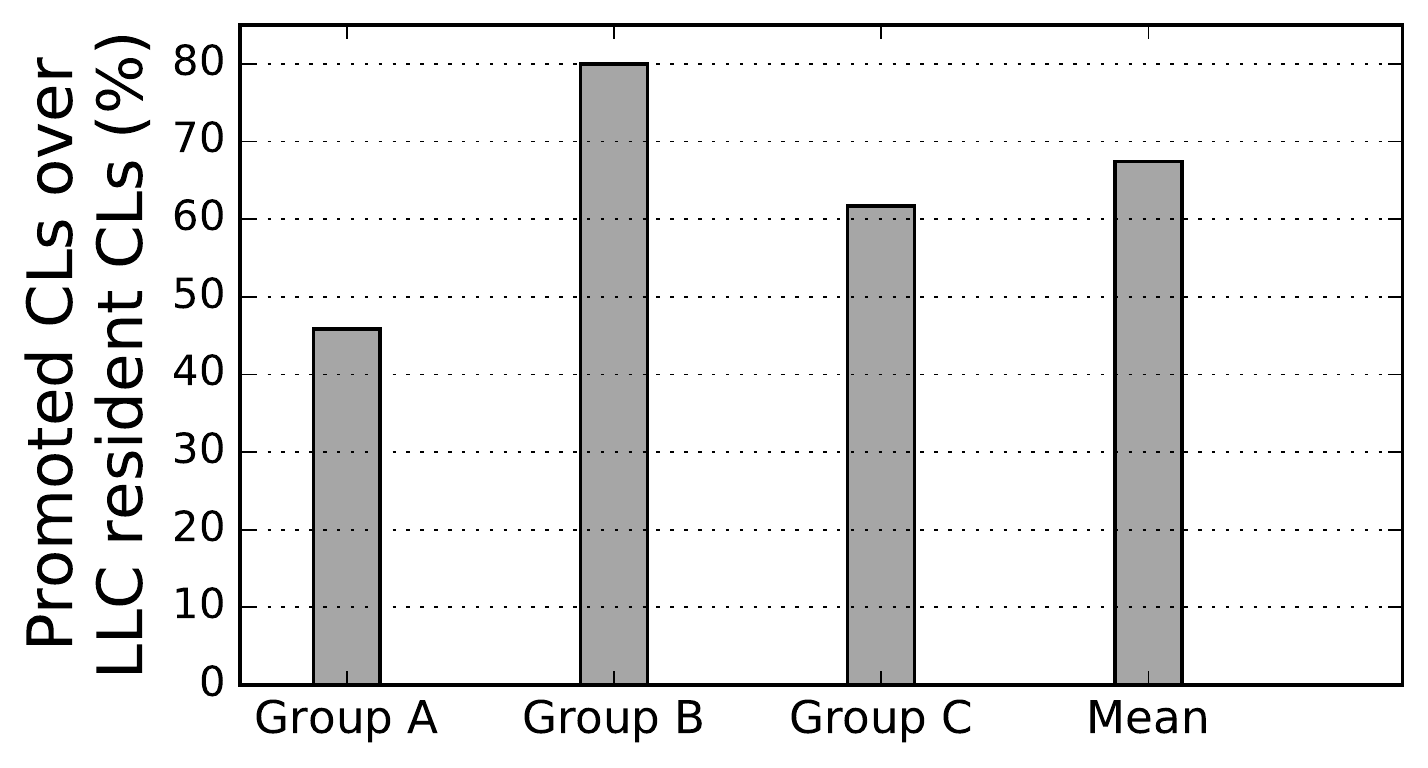}}
    \vspace{-2.mm}
    \caption{Characterizing PB use: (a) percentage 
of cache lines promoted into PBs that are actually accessed from the PBs, and (b) percentage 
of the
LLC-resident cache lines (CLs) of a page that are promoted to a PB in a PTR.}
    \label{fig:coverage}
    \vspace{-5.6mm}
\end{figure}

We now quantify the coverage of PTRs to the LLC in Figures \ref{fig:pbhits_act} and \ref{fig:acts_per_hint}.\footnote{Group A, Group B, and Group C are the mean of the SPEC CPU\textsuperscript{\textregistered} 2017, SPEC CPU\textsuperscript{\textregistered} 2006, and Coral benchmarks of Table \ref{table:workloads}, respectively.}
Figure \ref{fig:pbhits_act} shows what percentage 
of the cache lines promoted into PBs are actually accessed from the PBs.
We see that, on average, CLRs reference 51.1\% of the cache lines promoted to the PBs. \arch{} attains this high number by
adopting a PB replacement algorithm that favors the victimization
of PBs with few lines or a  low number of hits.
Note that when we promote cache lines into a PB, we do not pollute 
the LLC or L2. This is because the PBs simply hold a copy of the data present in the NVM data array.

Figure \ref{fig:acts_per_hint} shows what percentage of the
LLC-resident cache lines of a page are promoted to a PB in a 
PTR. This number is not 100\% for two reasons. First, for a given page,
some of the lines 
from different regions in
the same physical row may conflict with each other,
and cannot all be promoted to the PB. Second, if the number of 
LLC-resident lines is less than a threshold, \arch{} does not
promote the page. On average, \arch{} promotes 68\% of the 
LLC-resident cache lines to a PB---or about 26 lines.

\subsection{Alternative \arch{} Design}
\vspace{-0.2mm}

To further highlight the benefits of \arch{}, we evaluate  a scheme that fetches NVM-resident cache lines to the L2 cache instead of to the PBs, using the same trigger as \arch{}. For this experiment, we keep the data layout we introduced for the NVM cache, so that we can identify the cache lines of a page with a single read operation. Moreover, as an optimization, we make sure that the L2 cache always prioritizes read requests from the core over prefetches from 
the LLC to the L2 cache.
In addition, when the L2 MSHR entries are heavily utilized (i.e., $\sim$90\%), we drop outstanding LLC-to-L2 prefetches.

We find that this design is not competitive with \arch{}: 
on average, it is 19.8\% slower than \arch{} and increases the writes to NVM by 183\%. 
This is because bulk prefetches from LLC to L2 saturate the interconnect, causing core requests to stall while arbitrating for the bus. Moreover, fetching many lines to L2 causes L2 thrashing, which in turn increases L2 misses. 
This is especially the case for benchmarks with high L2 MPKI such as XSBench. This benchmark takes $\sim$90\% longer to complete with the new design than with \arch{} because of the increased traffic between the L2 and the LLC. 
Only benchmarks with a small L2 MPKI and a high PB hit ratio, such as 450.soplex and 437.leslie3d, can benefit from this design,
and attain a performance that is comparable to 
\arch{}'s.

Similarly, an aggressive L2 prefetcher that tries to prefetch the same cache lines as \arch{} faces the same performance bottleneck. Further, if the \arch{} LLC data layout is not used, read requests from the core suffer from the low LLC read bandwidth 
resulting from the non-pipelined latency of  NVM data array accesses.

\subsection{Sensitivity Analysis}
\begin{figure}[h]
    \captionsetup[subfloat]{captionskip=-0mm}
    \centering
    \subfloat[]{\label{fig:size_sensitivity_miss_rate}\includegraphics[width=0.515\columnwidth]{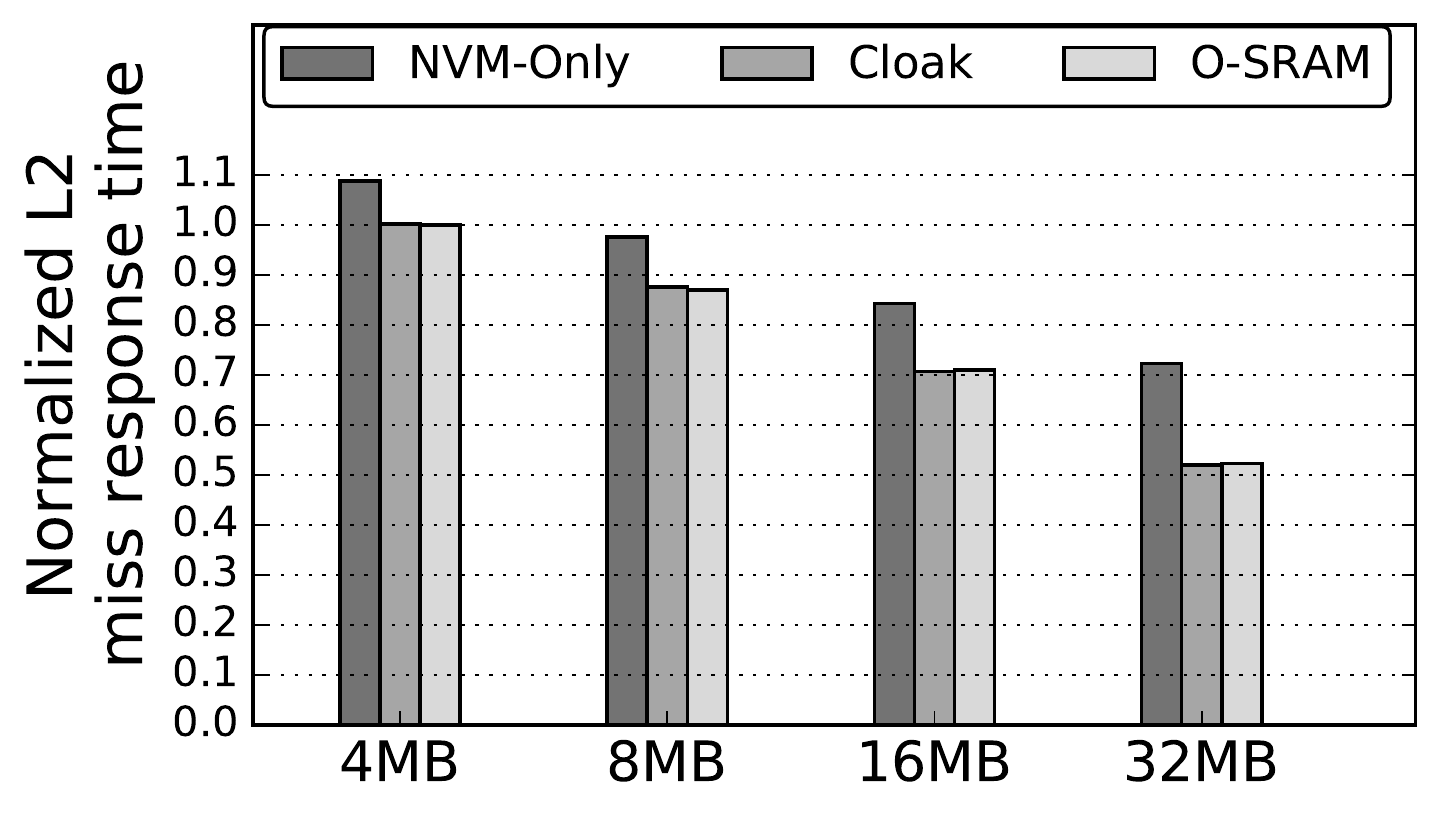}}
    \subfloat[]{\label{fig:size_sensitivity_perf}\includegraphics[width=0.49\columnwidth]{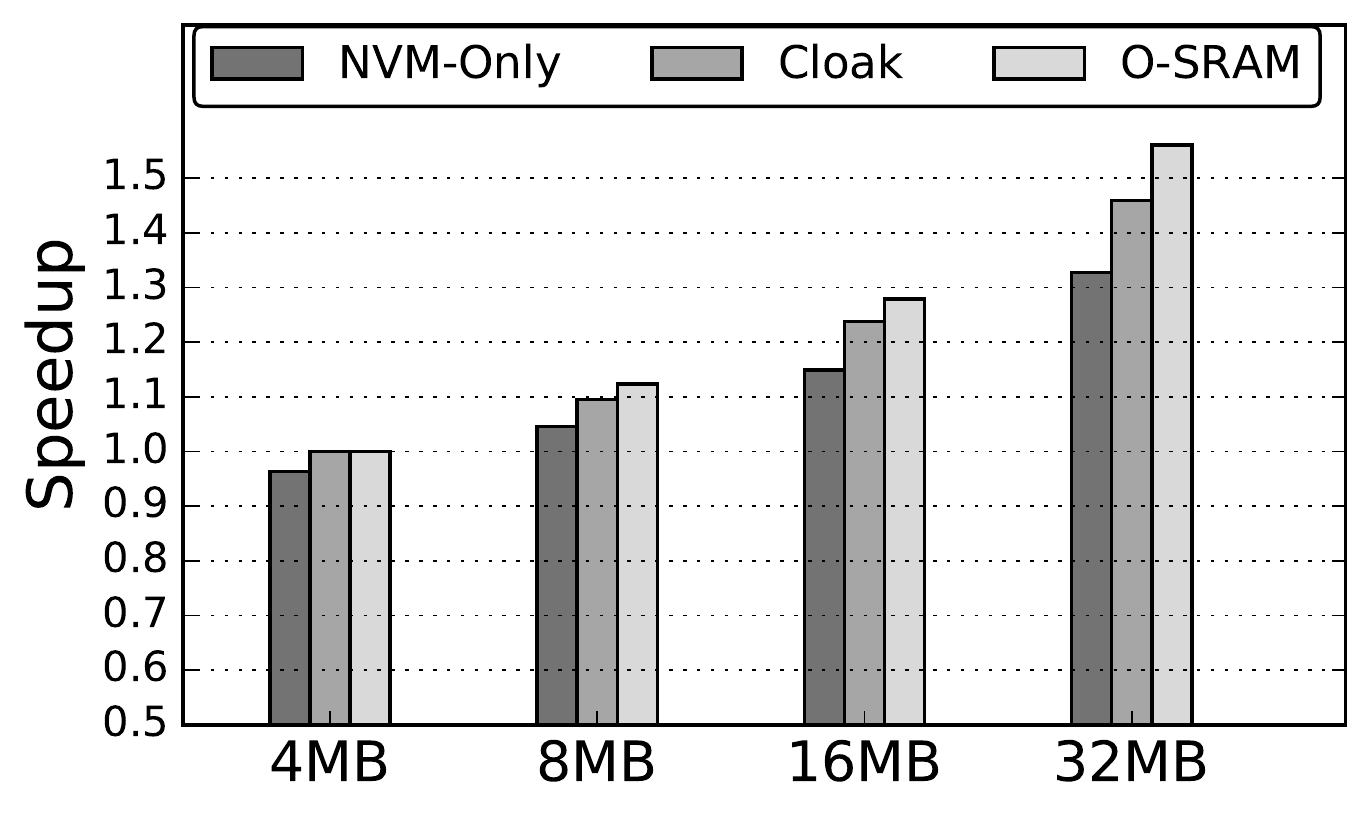}}
    \vspace{-1.9mm}
    \caption{Sensitivity analysis of different LLC sizes per core over Baseline with
    an SRAM-based LLC of 4MB per core:  (a) Normalized L2 miss response time and (b) speedup.}
    \label{fig:size_sensitivity}
    \vspace{-2.mm}
\end{figure}

\vspace{-0.2mm}
Finally, we perform two sensitivity analyses. First,
we examine  the sensitivity of \arch{} to the LLC cache size,
which is the primary parameter dictating the LLC hit/miss rate.
Figures \ref{fig:size_sensitivity_miss_rate} and
\ref{fig:size_sensitivity_perf} show the
average L2 miss response time and the average speedup, 
respectively, across all benchmarks, as the size of the LLC cache increases from 4MB to 32MB per core.
All results are normalized to Baseline, which has an SRAM-based 
LLC with  4MB per core.

Figure \ref{fig:size_sensitivity_miss_rate} shows that 
the relative L2 miss response time drops with the increase
in LLC size for all the schemes. \arch{} has lower L2 miss response time than NVM-Only for all configurations. It has
practically the same L2 miss response time as
O-SRAM because the PBs provide even faster access than
a larger SRAM LLC slice due to their smaller routing overhead.

Figure \ref{fig:size_sensitivity_perf} shows that 
the speedup of all the schemes increases with the LLC size.
This is because of the increasingly lower LLC miss rate.
For all LLC sizes, \arch{} delivers higher speedups than
NVM-Only and lower speedups than O-SRAM. Interestingly,
\arch{} can tolerate the higher read latency of NVM and achieve equal performance as the Baseline already with a 4MB LLC.

\begin{figure}[h]
    \captionsetup[subfloat]{captionskip=-0mm}
    \centering
    \vspace{-6.2mm}
    \subfloat[]{\label{fig:latency_sensitivity_L3MemTime}\includegraphics[width=0.515\columnwidth]{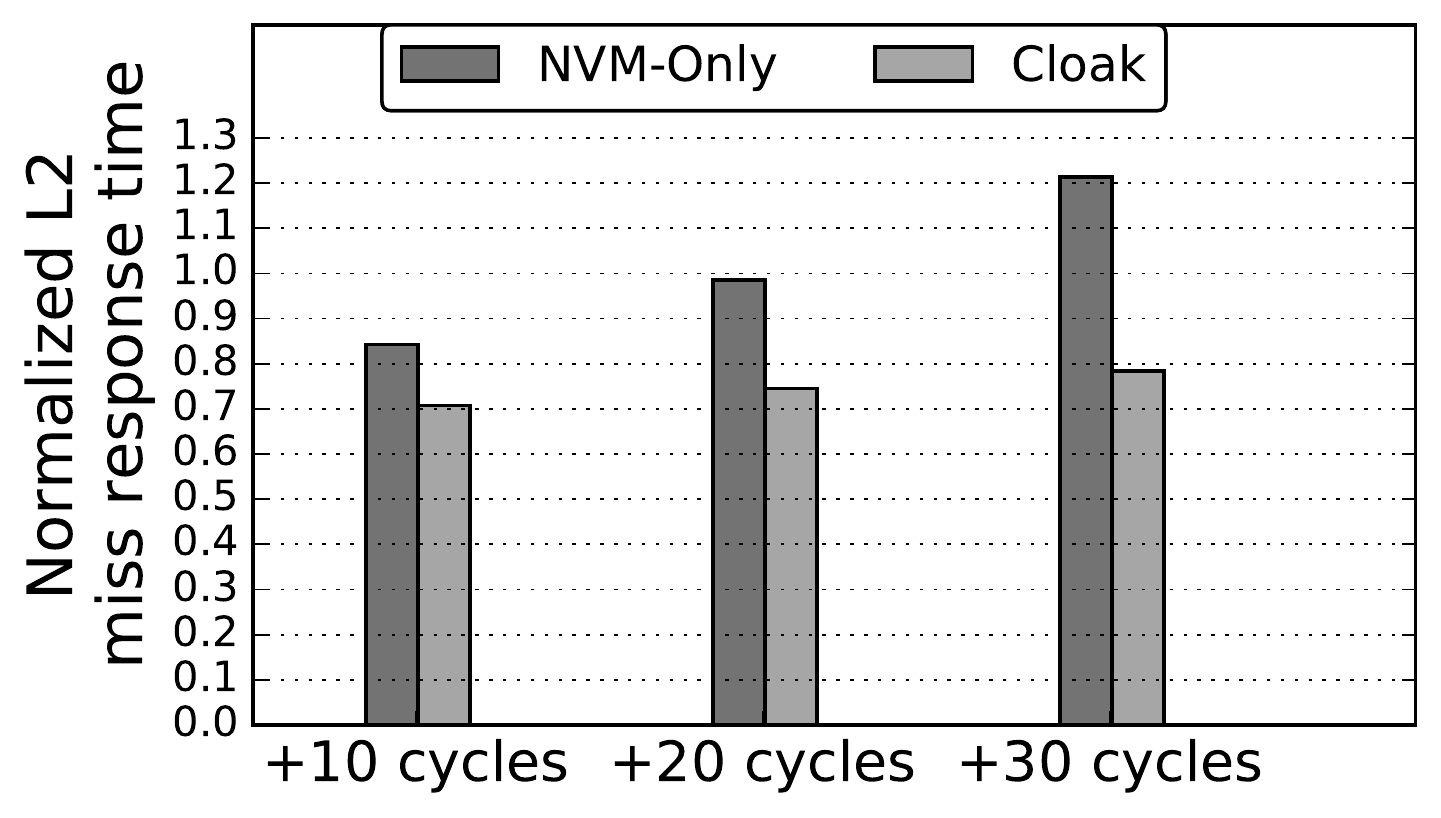}}
    \subfloat[]{\label{fig:latency_sensitivity_perf}\includegraphics[width=0.49\columnwidth]{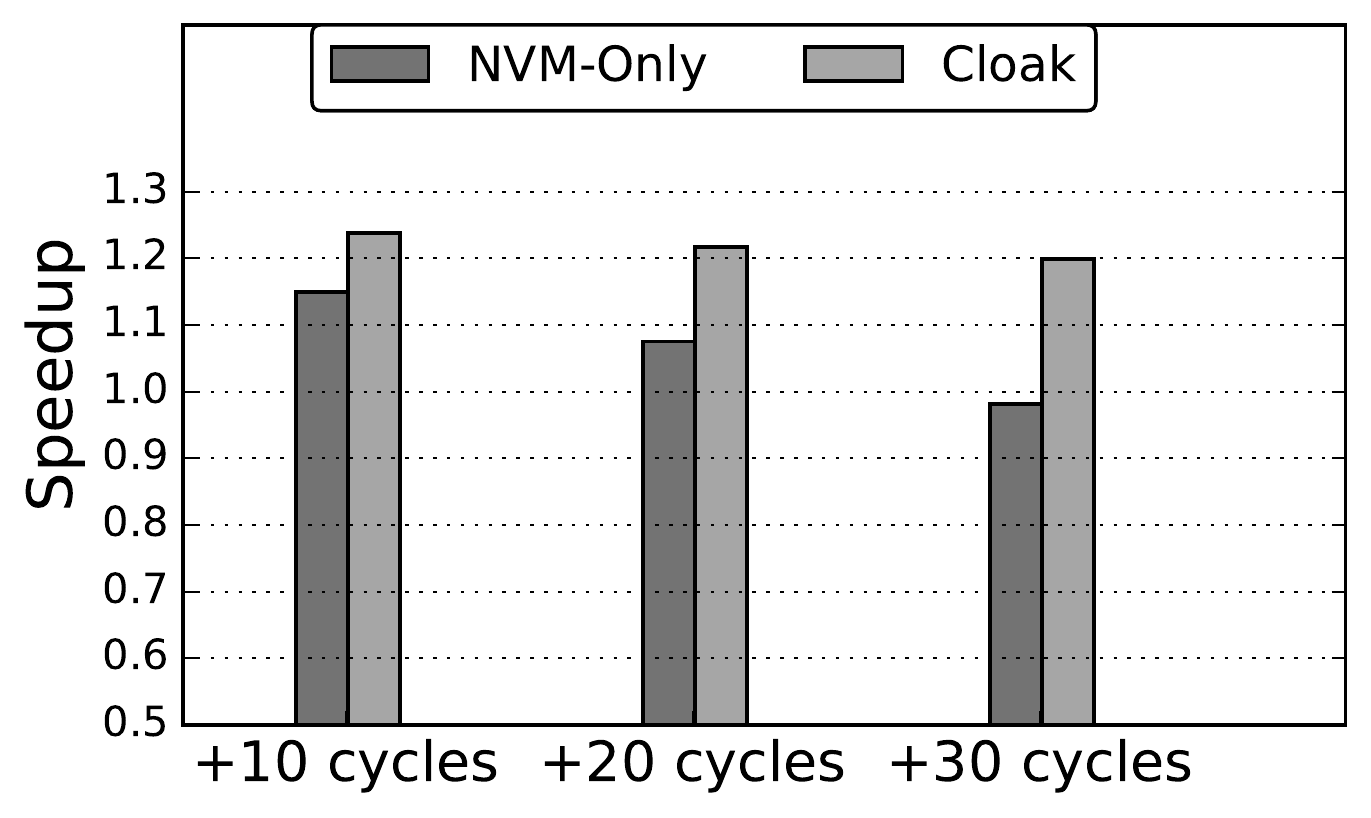}}
    \vspace{-1.9mm}
    \caption{Sensitivity analysis of different LLC read latencies over   Baseline with an SRAM-based LLC of 4MB per core:  (a) Normalized L2 miss response time and (b) speedup.}
    \label{fig:latency_sensitivity}
    \vspace{-2.1mm}
\end{figure}

We also analyze the effects of increasing the read latency of  NVM LLC caches, while keeping the cache size at 16MB per core. 
Figure \ref{fig:latency_sensitivity_L3MemTime} and
Figure \ref{fig:latency_sensitivity_perf} show the
average L2 miss response time and the average speedup, 
respectively, across all benchmarks, as the 
LLC read latency is increased. We increase the latency
by lengthening the NVM-based LLC data array read latency by 10, 20 and 30 cycles over 
the SRAM baseline. The configuration with +10 cycles represents the NVM cache we simulated in all of our prior experiments.
All results are normalized to Baseline, which has an SRAM-based 
LLC of 4MB per core.
The three designs represent an STT-RAM with a minimum read latency of $\sim$3 ns, $\sim$6 ns and 
$\sim$9 ns, respectively.

As the NVM LLC read latency increases, 
the relative L2 miss response time  increases,  and
the  speedup decreases. These trends occur for both
NVM-Only and \arch{}, although they are less prominent
for \arch{}. In all cases, \arch{} has a lower
L2 miss response time and a higher speedup than
NVM-Only. This is because the PBs can tolerate part of the higher NVM array read latency. Even with an NVM with 
30 cycles over 
the SRAM baseline, \arch{} is faster than Baseline.

\section{Other Related Work} \label{other_related}

\noindent {\bf Page Caches. }
Prior work has looked into the use of die-stacked eDRAM as large LLCs \cite{alloy_cache, unison_cache, footprint_cache, chop_dram_cache, loh_hill_dram_cache, tagless_dram_cache, tagless_dram_cache2, 3d_stacked}.
eDRAM-based caches are typically organized in pages instead of blocks to avoid massive tag storage. They are called {\em Page Caches}. 
When a request reaches a page cache and the page is not cached, the whole or a subset of the page \cite{footprint_cache, unison_cache} is brought from main memory, generating off-chip traffic. \arch{} does not generate any off-chip traffic.
The capacity of page caches is underutilized, since a page allocates cache space even for lines that are not fetched. This reduces cache capacity. In addition, page caches add extra overhead to keep track of a page's useful footprint. \arch{} does not
sacrifice any LLC capacity and does not need to track any footprint. Instead, it brings the LLC-resident lines into the PBs.

Page caches cannot be easily designed as victim or non-inclusive LLC
caches---e.g., storing a victim line requires the allocation of space for the whole page. Instead, \arch{} can be easily integrated with
LLCs with different inclusion properties.
Finally, if page caches are employed as an extra cache level (e.g., L4), \arch{} can still replace the SRAM-based shared L3 in such a design.

\noindent {\bf Techniques to Hide High Latency. }
To hide the increased latency of NVM caches, in addition to the
advanced techniques discussed in Section \ref{motivation}, one can use
conventional techniques such as prefetching and dead block elimination \cite{dasca}.
These proposals are orthogonal to \arch{} and can be used in conjunction with it.
However, LLC prefetchers incur increased complexity and 
can saturate memory bandwidth when using NVM caches \cite{stt_prefetch, oap, oscar}. 
The advantage of using the address translation hardware to make 
early decisions 
has been demonstrated before for page walks. Specifically,
TEMPO \cite{abhishek} uses  PTE page walk requests that miss in the cache hierarchy to prefetch from main memory to the LLC, the cache line that caused the page walk. 
PageSeer \cite{pageseer} uses page walk information to swap pages in a DRAM-NVM hybrid main memory system.
\arch{} is different from these approaches. First, it has a different
target (i.e., the set of LLC-resident cache lines of a page); 
second, it uses a different trigger (i.e., TLB miss on a page
used in the past).

\vspace{-0.5mm}
\section{Conclusion}

This paper presented \arch{}, a novel, low cost NVM LLC architecture that 
uses small SRAM-based page buffers to tolerate
the higher and non-pipelined latency of NVM reads. 
An L1 TLB miss on certain pages triggers the data transfer of LLC-resident lines belonging to the page from the NVM LLC to the page buffers. 
The buffers will service 
subsequent requests for this page, and use a novel replacement algorithm
to achieve high performance and low energy consumption.
\arch{} effectively hides the higher latency of NVM reads.
On average, \arch{} outperformed an SRAM LLC by 23.8\% and
an NVM-only LLC by 8.9\%---in both cases, with negligible additional area. Further,
the \(ED^2\) of \arch{} was 39.9\% and 17.5\% lower, respectively, than these two designs.


\bibliographystyle{IEEEtranS}
\bibliography{refs}

\end{document}